\documentclass[twocolumn]{aastex63}

\usepackage{amsmath}

\newcommand{\mdot}{\mbox{$\dot{M}$}}

\newcommand{\Rsun}{\mbox{$R_\odot$}}
\newcommand{\vinf}{\mbox{$v_\infty$}}

\newcounter{ion}

\definecolor{christi}{rgb}{0.0, 0.58, 0.71}

\definecolor{rico}{rgb}{0.0, 0.0, 0.60}

\shorttitle{Radio SEDs with Free-Free and Synchrotron Emission}
\shortauthors{Erba \& Ignace}

\begin{document}
\title{Radio Spectral Energy Distributions for Single Massive Star Winds with Free-Free and Synchrotron Emission}


\author{Christiana Erba}
\email{christi.erba@gmail.com}
\affil{Department of Physics and Astronomy, East Tennessee State University, Johnson City, TN 37663, USA }

\author{Richard Ignace}
%\email{ignace@etsu.edu}
\affil{Department of Physics and Astronomy, East Tennessee State University, Johnson City, TN 37663, USA }

%\altaffiltext{1}{Visiting Astronomer, Cerro Tololo Inter-American Observatory.
%CTIO is operated by AURA, Inc.\ under contract to the National Science
%\altaffiltext{2}{Society of Fellows, Harvard University.}
%\altaffiltext{3}{present address: Center for Astrophysics,
%    60 Garden Street, Cambridge, MA 02138}
%\altaffiltext{4}{Visiting Programmer, Space Telescope Science Institute}
%\altaffiltext{5}{Patron, Alonso's Bar and Grill}

%%%%%%%%%%%%%%%%%%%%%%%%%
\begin{abstract}    

The mass-loss rates from single massive stars are high enough to form radio photospheres at large distances from the stellar surface where the wind is optically thick to (thermal) free-free opacity. Here we calculate the far-infrared, millimeter, and radio band spectral energy distributions (SEDs) that can result from the combination of free-free processes and synchrotron emission, to explore the conditions for non-thermal SEDs. Simplifying assumptions are adopted in terms of scaling relations for the magnetic field strength and the spatial distribution of relativistic electrons. The wind is assumed to be spherically symmetric, and we consider the effect of Razin suppression on the synchrotron emission.  Under these conditions, long-wavelength SEDs with synchrotron emission can be either more steep or more shallow than the canonical asymptotic power-law SED from a non-magnetic wind. When non-thermal emission is present, the resultant SED shape is generally not a power-law; however, the variation in slope can change slowly with wavelength. Consequently, over a limited range of wavelengths, the SED can masquerade as approximately a power law. While most observed non-thermal long-wavelength spectra are associated with binarity, synchroton emission can have only a mild influence on single-star SEDs, requiring finer levels of wavelength sampling for detection of the effect.

\end{abstract}



%%%%%%%%%%%%%%%%%%%%%%%%%%%%%%%%%%%%%%%%%%%%%%%%%%

% Reminder: remove all \boldsymbol, \bf, and \mathbf before final submission

%\rico{Hi Rico, this is your color.}

\section{Introduction}

Massive stars are an important part of the story of cosmic evolution
\citep[e.g.,][]{2004ARA&A..42...79B,2010ApJ...724..341H,2014ARA&A..52..415M}
due to their luminous but short lifetimes \citep[e.g.,][]{2012ARA&A..50..107L}, their explosive endings \citep[e.g.,][]{2002RvMP...74.1015W}, the extreme
remnants that they produce \citep[e.g.,][]{2003ApJ...591..288H}, and their strong influence on galactic evolution \citep[e.g.,][]{2012MNRAS.421.3522H}. 
Our developing understanding of massive star evolution is informed by factors such as stellar rotation, magnetism, and mass-loss rates. Ongoing observational and theoretical investigations continue to refine our ability to measure and interpret these properties for massive stars.

Radio studies are a foundational
approach for determining the relatively high mass-loss rates ($\dot{M}$) from massive stars, which can be large enough to produce radio photospheres that form in the stellar wind 
\citep{1975A&A....39....1P,
    1975MNRAS.170...41W,
    1980ApJ...238..196A, 1981ApJ...250..645A,1986ApJ...303..239A,
    1989ApJ...340..518B, 
    1995ApJ...450..289L,1997ApJ...481..898L}.  
The radio flux (${\cal F}_\nu$) is typically attributed to thermal (free-free) emission, and is proportional to both the mass-loss rate of the star and the terminal velocity of the wind ($\vinf$) via ${\cal F}_\nu \propto (\mdot/\vinf)^{4/3}$ \citep[][]{1996ASPC...98..162L}. However, additional factors such as globally structured magnetic fields \citep[e.g.,][]{2004ApJ...600.1004O,2019MNRAS.489.3251D}, non-thermal emission from gyrosynchrotron processes \citep[e.g.,][]{1985ApJ...289..698W,1994Ap&SS.221..259C,2006A&A...452.1011V}, and  the existence of structured wind flows \citep[``clumping''; e.g.,][]{1997A&A...323..886B,1998A&A...333..956N,2006ApJ...637.1025F,2006A&A...454..625P,2016MNRAS.457.4123I}, have presented challenges for how radio measures can be used to infer $\dot{M}$ values. An improved understanding of how these processes affect the shape of the spectral energy distribution (SED), and thus interpretations of the stellar mass-loss rate, is therefore required. 

%In recent years, t
The discussion of non-thermal radio emission from massive stars has often focused on synchrotron emission in massive colliding wind binaries
\citep[CWB; e.g.,][]{1994A&A...291..805W,
    2006A&A...446.1001P,
    2012MNRAS.423.1562F,
    2014A&A...570A..10P,
    2015A&A...579A..99B,
    2017A&A...600A..47D}.
Electrons are accelerated at the shocks formed from the collision of highly supersonic wind flows, 
%It is still the case that 
and the {\em in situ} magnetism where synchrotron emission is generated derives from a stellar magnetic field. A tremendous advantage of a CWB system is the opportunity to observe cyclical variations in the non-thermal spectrum. This variability provides diagnostic leverage to produce globally self-consistent models for the wind properties, shock properties, magnetism, and the physics underlying particle acceleration \citep[e.g.,][]{2017A&A...608A..69B,2018BSRSL..87..185D}.

Although CWB systems have considerable potential to describe the non-thermal emission from massive star binaries, they cannot be used to fully characterize the non-thermal emission from (purportedly) single early-type stars. In recent years, an increasing number of these have been detected, in no small part due to the successes of the spectropolarimetric surveys that have provided direct measurements of large-scale, nearly dipoler\footnote{The assumption of a (nearly) dipolar magnetic topology is generally appropriate \citep[e.g.,][]{2009ARA&A..47..333D}. However, there are some notable exceptions, including $\tau$ Sco \citep[HD 149438, B0.2 V;][]{Donati2006b}, HD 37776 \citep[B2V;][]{Kochukhov2011}, and $\alpha^2$ CVn \citep[B9p;][]{Silvester2014}, which have fields with significantly more complex magnetic topologies.} 
surface magnetic fields in massive stars \citep[e.g., MiMeS, BOB;][]{
    2014A&A...567A..28A,
    2015A&A...582A..45F,
    2015IAUS..307..342M,
    2016MNRAS.456....2W,
    2017MNRAS.465.2432G}.
The magnetic fields of O- and B-type stars channel the stellar wind into a structurally complex magnetosphere, which can provide the conditions necessary for synchrotron emission \citep[e.g.,][]{
2004A&A...418..593T,
2006A&A...458..831L,
2021MNRAS.507.1979L,
2022arXiv220105512S}.
Non-thermal radio spectra of single, magnetic massive stars have been observed in several instances \citep[e.g.,][]{2017MNRAS.465.2160K,2017MNRAS.467.2820L,2018MNRAS.476..562L}. 
%However, stellar magnetism is not a prerequisite, since non-thermal radio emission has also been observed in a few non-magnetic, purportedly single, early-type stars \citep[e.g.,][]{2000MNRAS.319.1005D}.

\citet{1985ApJ...289..698W} published a seminal paper exploring the modeling and application of radio synchrotron emission for massive
star winds. The work was later expanded by \citet{1991ApJ...366..512C,
    1992ApJ...387L..81W,
    1994Ap&SS.221..259C,
    1994Ap&SS.221..295W}, 
both in terms of including additional physical processes (e.g., inverse Compton cooling), and for consideration of very high-energy emissions. Two developments emerged from this series. First, it appeared that the majority of sources displaying non-thermal radio emission were in binaries, hence the increased focus on modeling of colliding wind binary systems \citep[e.g.,][]{2003A&A...409..217D,2021MNRAS.504.4204P}.
Second was the recognition that there are challenges for understanding how relativistic electrons can survive and/or be energized at large radii in the wind \citep{2004A&A...418..717V,2005A&A...433..313V,
2006A&A...452.1011V}. While these challenges remain, non-thermal radio emission has been observed in a few non-magnetic, purportedly single, early-type stars \citep[e.g.,][]{2000MNRAS.319.1005D}. There is thus a need to further develop the theoretical description of how free-free and non-thermal emission jointly affect the SED, in order to provide benchmarks for these data.

%\christi{Although the theoretical description of how synchrotron emission is produced is still being refined, thermal plus (presumably) synchrotron emission has been observed in a number of recent cases. List some WR stars. List some magnetic stars. Any other single-star cases we can point to?The need to develop further constraints on how synch effects the SED is therefore pressing, in order to provide benchmarks for these data.}

In this paper, we present several new models of far-infrared, millimeter, and radio band spectral energy distributions (SEDs) resulting from the combination of thermal (free-free) and non-thermal (synchrotron) emissions. Section \ref{sec:models} develops the model components.  In particular, our goal is to examine general trends for single-star winds, and to this end, we consider spherical winds and adopt scaling relations for the magnetic field strength and the density of high-energy electrons. Section \ref{sec:seds} reports the results of our analysis. Finally, in Section \ref{sec:conc}, we discuss these results in light of the current data and  interpretation of radio SEDs for massive-star winds.

%%%%%%%%%%%%%%%%%%%%%%%%%
\section{Modeling Radio SEDs}		
\label{sec:models}

%%%%%%%%%%%%%%%%%%%%%%%%%
\subsection{Thermal Free-Free Emission}
\label{sec:free-free}

Our model is based on a smooth, spherically symmetric wind. While it is clear that magnetism can alter the wind density and vector flow from a spherical geometry, a spherical wind is appropriate for a qualitative exploration of  
%exploring the qualitative behavior  
how synchrotron emission can influence radio SED shapes.  Additionally, unlike numerically intensive MHD simulations and detailed radiative transfer, semi-analytic approaches afforded by simplifying assumptions (such as a spherical wind density) allows for a rapid exploration of the parameter space.

The emission coefficient for thermal free-free processes, expressed in terms of the power per unit frequency per unit volume per steradian, is written as \citep[e.g.,][]{1986rpa..book.....R}
%\begin{linenomath*}
\begin{equation}
j_{\nu}^{\rm ff} = 5.45\times 10^{-39} \, Z_{\rm i}^2 \, n_{\rm e} n_{\rm i} \, T^{-1/2} \, g_{\nu}^{\rm ff} \, e^{-h\nu/kT},
\end{equation}
%\end{linenomath*}
where $Z_{\rm i}$ is the root-mean-square charge of the ion, $T$ is the gas temperature, $n_{\rm e}$ and $n_{\rm i}$ are respectively the free electron and ion densities, $\nu$ is the observation frequency, and $g_{\nu}^{\rm ff}$ is the free-free Gaunt factor \citep[e.g.,][]{1962RvMP...34..507B, 1988ApJ...327..477H}. In the exponential, $h$ and $k$ are the usual Planck and Boltzmann constants. The numerical constant has been evaluated in {\it cgs} units.

In the Rayleigh-Jeans limit appropriate at long wavelengths where $h\nu \ll kT$, the expression for the emission coefficient becomes

%\begin{linenomath*}
\begin{equation}
j_\nu^{\rm ff} \approx 
    5.45\times 10^{-14} \, 
    Z_{\rm i}^2 \, 
    \left( \frac{n_{\rm e} n_{\rm i}} {10^{26}} \right)
    \left(\frac{10^4}{T}\right)^{1/2} \, g_{\nu}^{\rm ff},
\end{equation}
%%\end{linenomath*}
where we have scaled the temperature and ion number densities by fiducial values similar to those found in massive star winds.

The absorption coefficient, written here as the product of the opacity coefficient and the mass density, is given by \citep[e.g.,][]{1986rpa..book.....R}
%\christi{units!! This is per cm} 
%\begin{linenomath*}
\begin{equation}
\label{eq:alphanu}
(\kappa_\nu\,\rho)^{\rm ff} = 3.7\times 10^8 \, Z_{\rm i}^2 \, n_{\rm e} n_{\rm i} \, T^{-1/2} \, \nu^{-3} \, g_{\nu}^{\rm ff} \left(1-e^{-h\nu/kT}\right), 
\end{equation}
%\end{linenomath*}
which in the Rayleigh-Jeans limit is approximated as
%\begin{linenomath*}
\begin{equation}
\label{eq:alphanu_approx}
(\kappa_\nu\,\rho)^{\rm ff} \approx 
0.0020 \, Z_{\rm i}^2\,\frac{n_{\rm e} n_{\rm i}}{10^{26}}\,
\left(\frac{10^4}{T}\right)^{3/2} \,\lambda^2 \,g_{\nu}^{\rm ff}.
\end{equation}
%\end{linenomath*}
Note that we now express the absorption coefficient in terms of the observation wavelength $\lambda$. The source function for free-free radiation is therefore

%\begin{linenomath*}
\begin{eqnarray}
S_{\nu}^{\rm ff} 
    & = &  
    \left(\frac{j_\nu}{\kappa_\nu\rho}\right)^{\rm ff}
    = B_\nu 
    \nonumber \\
	& \approx & 
	\frac{2kT}{\lambda^2} 
	= 2.8\times 10^{-12} \, \lambda^{-2} \left(\frac{T}{10^4}\right).
\end{eqnarray}
%\end{linenomath*}

We can estimate the extent of the radial photosphere \citep[][]{1977ApJ...212..488C} assuming the wind is optically thick to the free-free opacity (eq.~[\ref{eq:alphanu_approx}]), with speed $v(r) \approx \vinf$.
The optical depth of a location in the wind is given by the integral of the free-free opacity along the observer's line-of-sight,
%\begin{linenomath*}
\begin{equation}
\label{eq:opd_ff}
\tau_{\nu}^{\rm ff} = \int (\kappa_\nu\rho)^{\rm ff}\,dz,
\end{equation}
%\end{linenomath*}
where $z = \varpi \cot \theta$, $\varpi$ is the impact parameter for a ray through the wind, and $\theta$ gives the angle between the radial direction and the observer's line-of-sight.
%$\varpi=\sin\theta/u$ is the impact parameter for a ray through the wind

We introduce $R_\nu$ as the radius at which an optical depth of unity is achieved, defined through
%\begin{linenomath*}
\begin{equation}
\label{eq:tau_unity}
\tau_{\nu}^{\rm ff} = \int_{R_\nu}^\infty \, (\kappa_\nu\rho)^{\rm ff} \, dr = 1.
\end{equation}
%\end{linenomath*}
For an isothermal wind with $T=10^4$~K, setting $Z_{\rm i} =1$, along with $n_{\rm e} = n_{\rm i}$, the effective radius $R_\nu$, expressed in terms of the stellar radius $R_{\ast}$, is

%\begin{linenomath*}
\begin{equation}
\label{eq:rnu_ff}
\frac{R_\nu}{R_\ast} \approx
    405 \left(\frac{\lambda}{\lambda_0}\right)^{2/3}
    \left(\frac{n_0}{10^{13}}\right)^{2/3}
	\left(\frac{R_\ast}{10^{11}}\right)^{1/3}\,(g_\nu^{\rm ff})^{1/3},
\end{equation}
%\end{linenomath*}
Here, we scale the wavelength by a fiducial value of $\lambda_0~=~1~{\rm cm}$, and the stellar radius by $10^{11}~{\rm cm}~\sim~1.5~\Rsun$. The number density scale constant $n_0$,
%\begin{linenomath*}
\begin{equation}
    \label{eq:n0}
    n_0 = \frac{\dot{M}}{4\pi \,
    \mu_{\rm e}\,m_{\rm H} R_\ast^2 \, v_\infty},
\end{equation}
%\end{linenomath*}
is derived from mass continuity (assuming spherical symmetry), where $\dot{M}$ is the mass-loss rate, $\vinf$ is the terminal velocity of the wind, and $\mu_{\rm e}$
is the mean molecular weight per free electron. With $n_{\rm e} = n_{\rm i}$,
a wind of entirely ionized hydrogen would have $\mu_{\rm e}=1$, whereas a wind
of entirely singly-ionized helium would have $\mu_{\rm e}=4$. 

Equation (\ref{eq:rnu_ff}) shows that the extent of the radio photosphere increases with wavelength, albeit the increase is more shallow than that from a linear dependence on $\lambda$. The radio photosphere at $\lambda_0 = 1~{\rm cm}$, assuming $n_0 = 10^{13}$~cm$^{-3}$, is extensive\footnote{A wind density scale of $n_0 = 10^{13}$~cm$^{-3}$ is applicable to WR stars, which have a very dense wind environment. For OB supergiants, a wind density scale of $n_0 = 10^{10}$~cm$^{-3}$ would be more appropriate. See also Appendix~\ref{app:appC}.}; 
however, ion abundances, the temperature and ionization state of the gas, and the stellar radius will realistically alter the value of $R_{\nu}$, even assuming that $n_0$ does not change. Furthermore, the presence of clumping \citep[structured wind flows believed to result from the intrinsic instabilities associated with the wind-driving physics, see e.g.,][]{1980ApJ...241..300L,1988ApJ...335..914O} in the wind would also extend the radio photosphere \citep{2016MNRAS.457.4123I}. 
%as \christi{define variable} ${\cal F}_{\rm cl}^{-1/3}$ \christi{is this the correct citation?} 

Radio SEDs purely from free-free opacity have been well-studied. \citet{1975MNRAS.170...41W} and \citet{1975A&A....39....1P} derived the canonical SED power-law result, with the radio flux ${\cal F}_\nu \propto \nu^{-0.6}$ for a spherical wind, at long ($\sim 1~{\rm cm}$) wavelengths for which the extent of the radio photosphere is much greater than the stellar radius. 
%\citet{1977ApJ...212..488C} considered power-law distributions for density and temperature in the RJ-Limit \christi{and concluded...}. 
%For spherical winds, \citet{1975MNRAS.170...41W} and \citet{1975A&A....39....1P} evaluated the radiative transfer, and for long wavelengths for which the radio photosphere is of much greater extent than the stellar radius, they arrived at the canonical SED power-law result that ${\cal F}_\nu \propto \nu^{-0.6}$.  
\citet{1982A&A...108..169S} found that under the assumption of  power-law distributions for the wind density and temperature, axisymmetric winds have SED slopes with the same power-law dependence of the frequency ($ \propto \nu^{-0.6}$) as from spherically symmetric winds (although the luminosity level is different from the spherical result).
%found that with \christi{under the assumption of a} power-law \christi{description of the} density and temperature, even  yield the same frequency dependence  of the SED slopes, and only the luminosity level that differs from the spherical result. 
This highlighted the idea that the SED slope is governed by the isophotal growth rate as a function of wavelength. Power-law SEDs result for isophotes whose shapes are invariant with wavelength. Ultimately, the expanding radio photosphere with wavelength provides an opportunity for mapping the wind density and geometry as a function of wavelength. For example, the canonical result of ${\cal F}_\nu \propto \nu^{-0.6}$, while not unique, can generally be taken as evidence of a spherically symmetric wind expanding at a constant speed.

Free-free radio SEDs have been evaluated for a variety of additional considerations. A change in the SED power-law slope can result from a change in the geometry of the circumstellar medium. The study of Be stars performed by \cite{2017A&A...601A..74K} provides a useful example. The authors found that disk truncation can lead to a SED slope that steepens toward ${\cal F}_{\nu} \propto \nu^{-2}$, which is suggestive of an as-yet-undetected binary companion. The influence of clumping for free-free radio SEDs has also been considered \citep[e.g.,][]{1981ApJ...250..645A, 1997A&A...323..886B, 2003ApJ...596..538I, 2016MNRAS.457.4123I}. 
Clumping affects the opacity, impacting the location of the effective radius $R_{\nu}$. Clumping also increases the emissivity, which scales as the square of density. The end result is that mass-loss rates derived from radio luminosities will be overestimated if clumping is not taken into account.

%%%%%%%%%%%%%%%%%%%%%%%%%
\subsection{Synchrotron Emission}
\label{sec:synch_emiss}

For a single electron, the specific luminosity of synchrotron emission is \citep[][]{1986rpa..book.....R,2003A&A...409..217D}
%\begin{linenomath*}
\begin{equation}
L_\nu = \sqrt{3}\,\frac{q^3\,B\,\sin\alpha}{m_{\rm e}\,c^2}\,
	F(\nu/\nu_{\rm c}),
\end{equation}
%\end{linenomath*}
where $B$ is the magnetic field strength, $m_e$ is the electron mass, $q$ is the electron charge, $\alpha$ is the pitch angle, and $F(x)$ gives the spectral slope as a function of frequency. The quantity $\nu_{\rm c}$ is the cut-off frequency, given by
%\begin{linenomath*}
\begin{equation}
\nu_{\rm c} = \frac{3\gamma^2\,q\,B\,\sin\alpha}{4\pi\,m_{\rm e}\,c},
\end{equation}
%\end{linenomath*}
where $\gamma$ is the Lorentz factor of the relativistic electrons. For an electron, $\nu_{\rm c} = 4.2~{\rm MHz}\,\gamma^2\,B\,\sin\alpha$, or in terms of wavelength, $\lambda_{\rm c} =
0.071~{\rm km}\,(\gamma^2\,B\,\sin\alpha)^{-1}$.

The synchrotron spectrum can be derived from a power-law distribution of free electrons. At a given location in the wind, the energy distribution of the electron number density can be represented as
%\begin{linenomath*}
\begin{equation}
\frac{dn_\gamma (r,E)}{dE}\,dE = C_E(r)\,E^{-p(r)}\,dE,
\end{equation}
%\end{linenomath*}
where $E$ is the energy, $C_E$ is a scale constant, and $p$ is the power-law exponent for the distribution. This can be recast in terms of the Lorentz factor for the relativistic electrons, with
%\begin{linenomath*}
\begin{equation}
\frac{dn_\gamma(r,\gamma)}{d\gamma}\,d\gamma = C_\gamma(r)\,
	\gamma^{-p(r)}\,d\gamma,
\end{equation}
%\end{linenomath*}
where $C_\gamma$ is the corresponding scale constant. Note that with respect to units, $C_\gamma \sim$ number density, but $C_E$ has units of energy that depend on the value of $p$. The ratio of the scale constants is thus
%\begin{linenomath*}
\begin{equation}
\label{eq:cgam_cE}
\frac{C_\gamma}{C_E} = (m_{\rm e}\,c^2)^{1-p}.
\end{equation}
%\end{linenomath*}
 The value of the scale constants $C_{\gamma}$ and $C_E$ can vary with
pitch angle \citep{1986rpa..book.....R}, but we ignore such pitch angle effects for this study.

The number density of relativistic electrons associated with the production of synchrotron emission is then
%\begin{linenomath*}
\begin{equation}
n_\gamma(r) = C_\gamma(r)\,\int_{\gamma_{\rm min}}^{\gamma_{\rm max}} \gamma^{-p}\,d\gamma.
\end{equation}
%\end{linenomath*}
The power-law index $p$ can vary with radius
\citep[e.g.,][]{2005A&A...433..313V}, but we choose\footnote{Other values could certainly be used: \citet{2004A&A...418..717V} suggest a steeper value with $p_0$ between 2 and 3. We adopt here the choice of $p_0=2$ to facilitate our quantitative examples.} a constant value of $p(r) = p_0 = 2$, which is commonly adopted for strong shocks \citep[see][and sources therein]{1985PhRvL..55.2735E,1993ApJ...402..271E}. 

If we assume $\gamma_{\rm min} \sim 1$ and $\gamma_{\rm max} \gg 1$ for all locations in
the wind, the number density of relativistic electrons becomes
%\begin{linenomath*}
\begin{equation}
\label{eq:cgamma}
n_\gamma(r) \approx C_\gamma(r) \equiv C_\ast\,\left(
	\frac{R_\ast}{r}\right)^{m+2},
\end{equation}
%\end{linenomath*}
where $m$ is a power-law exponent that is a free
parameter of our model. The case of $m=0$ corresponds to an inverse-square decline of density of high-energy elections $C_\gamma$, matching the scaling of the gas density with radius in the asymptotic constant expansion portion of the wind.

Simplifying for the specific case of $p_0=2$, we find that $C_E = C_\gamma \, m_{\rm e}\,c^2$. Assuming equipartition, the scale constant for the particle distribution is expressed as \citep{2003A&A...409..217D}
%\begin{linenomath*}
\begin{equation}
C_\gamma = \frac{u_{\rm rel}/m_{\rm e}\,c^2}{\ln\gamma_{\rm max}}\,\left(
	\frac{R_\ast}{r}\right)^{m+2},
\end{equation}
%\end{linenomath*}
where $u_{\rm rel}$ is the energy density of the relativistic particles, and $\gamma_{\rm max}$ is the largest value of $\gamma$ achieved by any particle.

The emission coefficient for synchrotron processes, expressed as power per unit frequency per unit volume per steradian, is \citep{1986rpa..book.....R}
%\begin{linenomath*}
\begin{eqnarray}
    j_{\nu}^{\rm s} & = & 
    \frac{\sqrt{3}q^3 B}{4 \pi m_{\rm e}\,c^2} \,
    \frac{C_\gamma}{p+1} \,
     \Gamma\left(\frac{p}{4} + \frac{19}{12}\right) \,
	\Gamma\left(\frac{p}{4}-\frac{1}{12}\right) \,
    \nonumber \\
	& \times &
	\left(\frac{2\pi\,m_{\rm e}
	\,c\,\nu}{3q B}\right)^{(1-p)/2}
	.
\end{eqnarray}
%\end{linenomath*}
where we choose $\alpha = 90^\circ$ for the pitch angle. 

The absorption coefficient for synchrotron processes is given by \citep{1986rpa..book.....R}
%\begin{linenomath*}
\begin{eqnarray}
    (\kappa_\nu\rho)^{\rm s} & = & 
    \frac{\sqrt{3}q^3}{8\pi\,m_{\rm e}} \,
	\left(\frac{3q}{2\pi\,m_{\rm e}^3\,c^5}\right)^{p/2}
	B^{(p+2)/2} \,
	\nu^{-(p+4)/2} \nonumber \\
	& \times & 
	C_E\,\Gamma\left(\frac{p}{4}+\frac{1}{6}\right)\,\Gamma\left(
	\frac{p}{4}+\frac{11}{6}\right) 
	.
\end{eqnarray}
%\end{linenomath*}
The source function for the synchrotron emission
is then
%\begin{linenomath*}
\begin{eqnarray}
    S_{\nu}^{\rm s} & = & 
    \frac{j_{\nu}^{\rm s}}{(\kappa_\nu\rho)^{\rm s}} \\
	& = & 
	\frac{2 \, m_{\rm e}^{p} \, c^{2p} \, \nu^{5/2}}{c^2} \,
	\left(\frac{2\pi\,m_{\rm e}\,c}{3qB}\right)^{1/2}
	\frac{C_\gamma}{C_E} \,
	\frac{\tilde{\Gamma}(p)}{p+1}
	,
\end{eqnarray}
%\end{linenomath*}
where
%\begin{linenomath*}
\begin{equation}
\tilde{\Gamma}(p) \equiv\frac{\Gamma(a_1)\,\Gamma(a_2)}{\Gamma(a_3)\,
	\Gamma(a_4)},
\end{equation}
%\end{linenomath*}
with
%\begin{linenomath*}
\begin{eqnarray}
a_1 & = & (3p+19)/12, \\
a_2 & = & (3p-1)/12,  \\
a_3 & = & (3p+2)/12, \\
a_4 & = & (3p+22)/12.
\end{eqnarray}
%\end{linenomath*}
Using $p_0=2$ and equation (\ref{eq:cgam_cE}), the expressions for the synchrotron emission coefficient, opacity, and source function reduce to
%\begin{linenomath*}
\begin{align}
    j_{\nu}^{\rm s} & =  
    2.3\times 10^{-15}\,\left(\frac{C_\gamma}{10^{10}}\right)\,B^{3/2}\,\lambda^{1/2}, \label{eq:jnu_p2} \\
    (\kappa_\nu\rho)^{\rm s} & =  5.1\times 10^{-11} \, \left(\frac{C_\gamma}{10^{10}}\right) \, B^2\,\lambda^3, 
    \label{eq:alphanu_p2}\\
    S_{\nu}^{\rm s} & =
    4.5\times 10^{-5} \, B^{-1/2} \, \lambda^{-5/2}
    .
\end{align}
%\end{linenomath*}
%where we have scaled $C_{\gamma}$ by a fiducial value of 1e10.

It is possible to develop an ``effective photosphere'' analysis for pure, optically thick synchrotron radiation, following the approach
of \citet{1977ApJ...212..488C}. Details on the full solution and the approximation of an effective photosphere are given in Appendix \ref{app:effphoto_synch}. With  equation~(\ref{eq:cgamma}), and a field strength that declines like a toroid\footnote{With flux freezing for highly ionized winds, rotation generally leads to a toroidal field topology \citep[e.g., ][]{1998ApJ...505..910I}.} with
%\begin{linenomath*}
\begin{equation}
\label{eq:synchB}
B(r) = B_\ast \left(\frac{R_\ast}{r}\right),
\end{equation}
%\end{linenomath*}
where the ``$\ast$'' notation indicates a value at the base of the stellar wind, the extent of the radio photosphere will then scale as $R_\nu \propto \lambda^{3/(3+m)}$. The corresponding specific radio luminosity becomes $L_\nu \propto S_\nu\,R^2_\nu \propto \lambda^{-5m/(6+2m)}$. 
%$L_\nu \propto \lambda^{-q(m)}$, where $q=5m/(6+2m)$.  
For a choice of $m=2$ in $C_\gamma$, the photospheric radius will grow as $R_\nu \sim \lambda^{0.6}$, close to the value for
free-free emission. However, the SED slope is somewhat steeper than the case of free-free, with $L_\nu \propto \lambda^{-1}$. We note that for large values of $m$, the radius of the synchrotron radio photosphere approaches a constant with wavelength, and the SED steepens to $L_\nu \sim \lambda^{-2.5}$. However,
there are other factors that influence the SED, such as Razin suppression, addressed below.

%%%%%%%%%%%%%%%%%%%%%%%%%
\subsection{The Razin Effect}
\label{sec:razin}

The Razin effect \citep[e.g.,][]{2003A&A...409..217D} refers to the suppression of synchrotron emission due to the refractive index of the plasma. It is dependent on the plasma frequency, and the frequency at which the effect becomes important is given by \citep[e.g.,][]{2004A&A...418..717V}:
%\begin{linenomath*}
\begin{equation}
\label{eq:razin_nu}
\nu_R = 20\,\frac{n_{\rm e}}{B}.
\end{equation}
%\end{linenomath*}
The suppression of the synchrotron emission scales as $\approx e^{-\nu_R/\nu} = e^{-\lambda/\lambda_R}$, where $\lambda_R = c / \nu_R$. At frequencies lower than $\nu_R$ (or wavelengths longer than $\lambda_R$), the synchrotron emission is strongly reduced.

It is useful to develop a scale of where in the wind the Razin effect becomes significant. To provide an estimate, we assume the electron density decreases as $r^{-2}$, and the field strength decreases as $r^{-1}$. With a wind density of $n_0 = 10^{13}$ cm$^{-3}$, and a stellar field strength
of $B_\ast = 100$~G, equation (\ref{eq:razin_nu}) gives $\lambda_R = 1$~cm at a radius of $r_R\approx
70R_\ast$. At larger radii, the Razin wavelength will be greater. Consequently, for an observation wavelength of 1~cm, the Razin effect will suppress synchrotron emission for $r<r_R$, but not for $r>r_R$.

%%%%%%%%%%%%%%%%%%%%%%%%%
\section{SEDs from Winds with Free-Free and Synchrotron Emission} 
\label{sec:seds}

%%%%%%%%%%
\begin{table}[t]
\begin{center}
\caption{Model parameters for the SEDs shown in Figure~\ref{fig:hybrid}. The dimensionless constant $a_0$ is used to parameterize the Razin wavelength, as $\lambda_{\rm R} = \lambda_{\rm R}^{0}/a_0$ (see also eq.~[\ref{eq:lam_r0}]). Choosing a value of $B_{\ast} =$~100~G, $K_0\sim100$ for typical WR star parameters, and $K_0\sim20$ for the typical parameters associated with an O supergiant.
\label{tab1}
}
\begin{tabular}{lcccc}
\hline \hline
$m$ & 0.0, 0.5, 1.0 \\
$K_0(m=0.0)$ & 1 \\
$K_0(m=0.5)$ & 100 \\
$K_0(m=1.0)$ & 30,000 \\
$a_0$ & 0 (no Razin), \\
      & 1, 3, 9, 27, 81 \\ \hline
\end{tabular}
\end{center}
\end{table}
%%%%%%%%%%
\begin{figure*}
\centering
\includegraphics[angle=0,width=1.95\columnwidth]{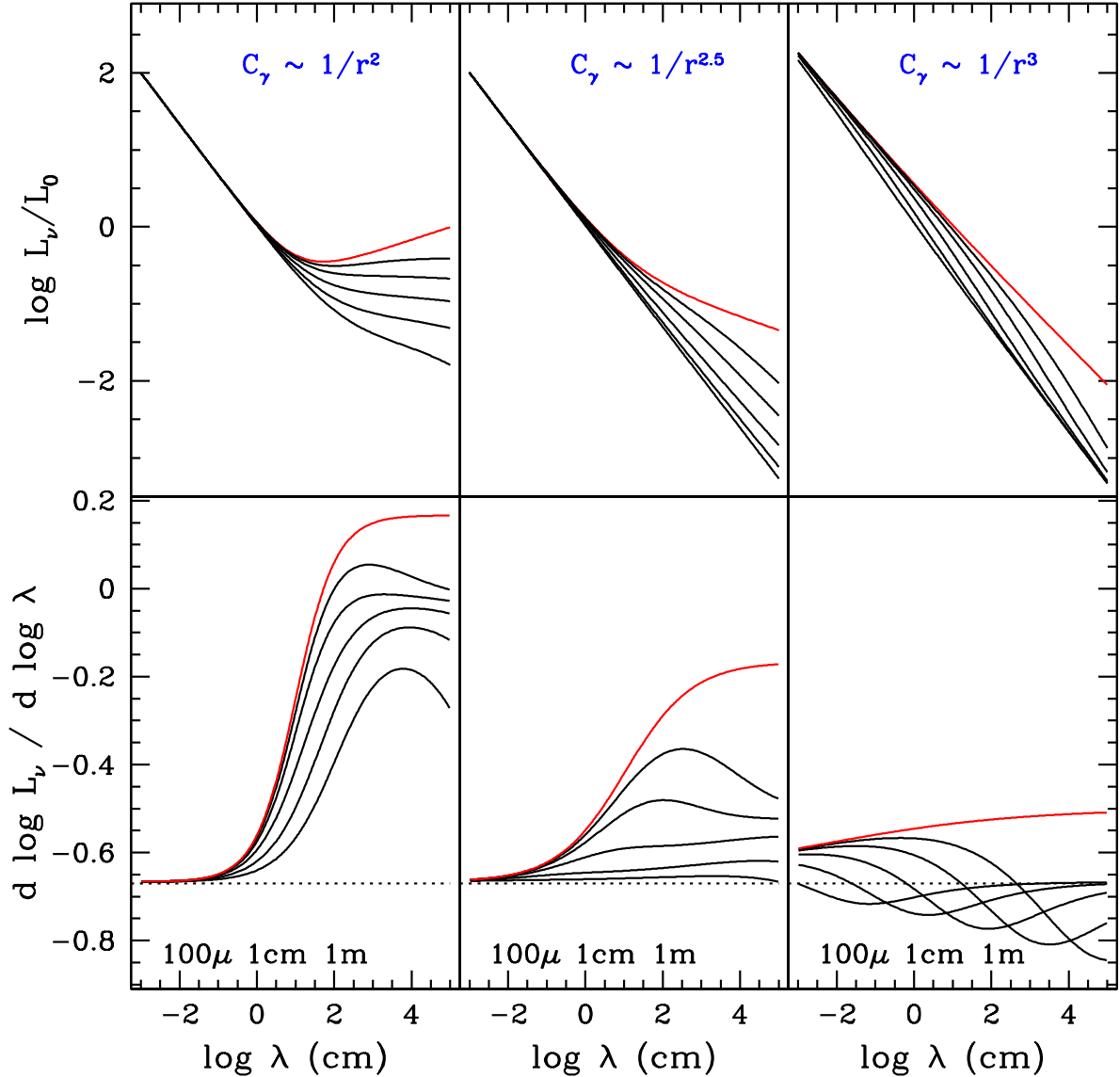}
\caption{The upper panels show model SEDs; the lower panels show logarithmic slopes of the SEDs with wavelength, representing the instantaneous power-law exponent. At short wavelengths where the slope is $-2/3$ (dotted line), the SED is dominated by the free-free emission. Deviations from that value signify the influence of synchrotron emission. The rows from left to right show different radial distributions of the relativistic electron population density, as represented by $C_\gamma$. The red curve in each case indicates a model that does not include the Razin effect (corresponding to the model parameter $a_0 = 0$). Each successive curve below is a model with a larger value of $a_0$, corresponding to increasingly more suppression of the synchrotron emission. The model parameters are provided in Table~\ref{tab1}. Consistent with the discussion in Section \ref{sec:synch_emiss}, we have used a power-law index of $p(r) = p_0 = 2$.}
\label{fig:hybrid} 
\end{figure*}
%%%%%%%%%%

We now consider an ionized and magnetized wind that is spherically symmetric, with both free-free and synchrotron opacities. The source function is
%\begin{linenomath*}
\begin{equation}
    S_\nu = 
    \frac{j_{\nu}^{\rm ff} + j_{\nu}^{\rm s}}
	{(\kappa_\nu\rho)^{\rm ff} + (\kappa_\nu\rho)^{\rm s}} .
\end{equation}
%\end{linenomath*}
Except for very long wavelengths or very large distances in the wind, it is reasonable to expect that the synchrotron opacity is considerably smaller than the free-free opacity (Appendix~\ref{app:effphoto_synch}), so that the source function reduces to
%\begin{linenomath*}
\begin{equation}
S_\nu = B_\nu(T) + \frac{j_{\nu}^{{\rm s}}}{(\kappa_\nu\rho)^{\rm ff}},
\end{equation}
%\end{linenomath*}
with the free-free opacity $(\kappa_\nu\rho)^{\rm ff}$ (eq.~[\ref{eq:alphanu_approx}]) as the sole contributor to the absorption coefficient. 

From equation (\ref{eq:opd_ff}), the optical depth of the thermal absorption is the line-of-sight integral of the free-free opacity, along a constant impact parameter. The expression is analytic, and integrates to
%\begin{linenomath*}
\begin{equation}
\label{eq:ff_opd}
    \tau_\nu^{\rm ff} = \frac{\tau_0}{2} \,
    u^3 \, g_{\nu}^{\rm ff} \, 
     \left(\frac{\lambda}{\lambda_0}\right)^2
    \left(\frac{\theta- \mu \sin\theta}{\sin^3\theta}\right),
\end{equation}
%\end{linenomath*}
where $\cos \theta = \mu$, $\sin\theta = \varpi u$, and $u = R_{\ast}/r$. The angle $\theta$ varies over a range of [0,~$\pi$], with $\theta < \pi/2$ describing the hemisphere of the observer.
For simplicity we have defined the constant
%\begin{linenomath*}
\begin{equation}
    \label{eq:tau0}
    \tau_0 =
    2 \times 10^{8} \, \frac{Z_{\rm i}^2}{\mu_{\rm i} \, \mu_{\rm e}}
    \left(\frac{R_{\ast}}{10^{11}}\right)
    \left( \frac{n_0}{10^{13}} \right)^2
    \left( \frac{10^4}{T} \right)^{3/2}
    .
\end{equation}
%\end{linenomath*}

With these quantities, and ignoring any radio emission from the star itself, the formal solution for a radio SED from an isothermal wind, is
%\begin{linenomath*}
\begin{eqnarray}
\label{eq:form_sol}
    L_\nu & = & 
    8\pi^2 R_{\ast}^{2} \left\{ B_\nu 
    \int_{0}^{\infty} \left[1-e^{-\tau_{\rm tot}(\varpi)} \right] \, \varpi\,d\varpi \right. 
    \nonumber \\
	& + & 
	\left. 
	\int_{0}^{\infty}
	\int_{0}^{\tau_{\rm tot}(\varpi)}
	\frac{j_{\nu}^{\rm s}} {(\kappa_\nu\rho)^{\rm ff}} \,
	e^{-\tau_{\nu}^{\rm ff}} \, \varpi\, d\tau_{\nu}^{\rm ff}\, d\varpi \right\},
\end{eqnarray}
%\end{linenomath*}
where 
%$\varpi=\sin\theta/u$ is the impact parameter for a ray through the wind, and 
$\tau_{\rm tot}=\tau_\nu^{\rm ff}(\varpi,\theta\to\pi)$, and we have ignored any contribution from the stellar disk.
Note that we ignore Razin suppression of the synchrotron emission in this expression, which will be discussed below.

The first integral in equation~\ref{eq:form_sol} can be evaluated analytically as
%\begin{linenomath*}
\begin{eqnarray}
    L_\nu &=& 
    \nonumber
    8\pi^2 R_{\ast}^{2} \; B_\nu 
    \int_{0}^{\infty} \left[1-e^{-\tau_{\rm tot}(\varpi)} \right] \, \varpi\, d\varpi
    \\
    \nonumber
    &=& 
    8\pi^2 R_{\ast}^{2} \; B_\nu 
    \int_{0}^{\infty} \left[1-e^{-\zeta \varpi^{-3} } \right] \, \varpi\,d\varpi
    \\
    &=&
    4\pi^2 R_{\ast}^{2} \; B_\nu \; \Gamma\left(\frac{1}{3}\right) 
    \left( \frac{\pi}{2} \; \tau_0 \; %\lambda^2 
     \left(\frac{\lambda}{\lambda_0}\right)^2
    \right)^{2/3},
\end{eqnarray}
%\end{linenomath*}
where in the second equality we have used
%\begin{linenomath*}
\begin{eqnarray}
    \tau_{\rm max} (\varpi) 
    &=&
    \nonumber
    \int_{- \infty}^{\infty} ( \kappa_{\nu} \rho )^{\rm ff} dz
    \\
    &=&
    \frac{\pi \; \tau_0 \; 
    %\lambda^2
    }{2 \; \varpi^3}
     \left(\frac{\lambda}{\lambda_0}\right)^2
    ,
\end{eqnarray}
%\end{linenomath*}
with $\zeta = \pi \tau_0 \lambda^2 / 2 \lambda_0^2$. Using the appropriate approximation for $B_\nu$ in the Rayleigh-Jeans limit, this expression can be recast as
%\begin{linenomath*}
\begin{equation}
    L_\nu = 
    L_0 \left( \frac{\lambda}{\lambda_0} \right)^{-2/3}
\end{equation}
%\end{linenomath*}
with
%\begin{linenomath*}
\begin{equation}
    L_0 =  
    1.35 \times 10^{18} 
    \left(\frac{Z_{\rm i}^{2}}{\mu_{\rm e} \, \mu_{\rm i}}\right)^{2/3}
    \left(\frac{n_{0}}{10^{13}}\right)^{4/3}
    \left(\frac{R_{\ast}}{10^{11}}\right)^{8/3}
\end{equation}
%\end{linenomath*}
as the scaling for the specific luminosity, with units of erg s$^{-1}$ Hz$^{-1}$.
For the sake of illustrative models, we ignore the wavelength dependence of the free-free factor and set $g_{\nu}^{\rm ff}=1$, yielding $L_{\nu} \propto \lambda^{-2/3}$. Inclusion of $g_{\nu}^{\rm ff}$ would produce the canonical $L_{\nu} \propto \lambda^{0.6}$ result.
 %\citep[e.g.,][]{1962RvMP...34..507B, 1988ApJ...327..477H}.

The second integral in equation (\ref{eq:form_sol}) can be recast as
%\begin{linenomath*}
\begin{equation}
    \label{eq:form_sol_2int}
    L_\nu = 8 \pi^2 \, R_{\ast}^3 \, j_0 \int_{-1}^{1}
    \int_{0}^{1} 
    %\lambda^{1/2} 
     \left(\frac{\lambda}{\lambda_0}\right)^{1/2}  
    \, u^{m-1/2} \, e^{-\tau_{\nu}^{\rm ff}} du~d\mu
\end{equation}
%\end{linenomath*}
where we have applied equation (\ref{eq:jnu_p2}), assuming $p=2$, and the constant $j_0 = 2.3 \times 10^{-25}~B_{\ast}^{3/2} C_{\ast}$. 
%We make the substitution $u = R_{\ast}/r$, and use $\mu = \cos \theta$, where $\theta$ gives the angle between the radial direction and the observer's line-of-sight.

For the case of $m=1/2$, for an optically thick wind ($\tau_0 \gg 1$), equation (\ref{eq:form_sol_2int}) can be solved analytically to express the wavelength dependence of the SED,
%\begin{linenomath*}
\begin{eqnarray}
    L_\nu & = & 8 \pi^2 \, R_{\ast}^3 \, j_0 \, %\lambda^{1/2}
     \left(\frac{\lambda}{\lambda_0}\right)^{1/2} 
    \int_{-1}^{1}
    \int_{0}^{\infty} \, e^{-\tau_{\nu}^{\rm ff}} du~d\mu
    \nonumber \\
    & = &
    %8 \pi^2 \, R_{\ast}^3 \, j_0 \,
    L_0 \, K_0 \, 
    %\lambda^{-1/6}
     \left(\frac{\lambda}{\lambda_0}\right)^{-1/6} 
    \left(\frac{2}{\tau_0} \right)^{1/3}
    \Gamma\left(\frac{4}{3}\right)\,\Lambda_0\,
    ,
\end{eqnarray}
%\end{linenomath*}
where
%\begin{linenomath*}
\begin{eqnarray}
    \Lambda_0 &=&
    \int_{-1}^{+1}\,\frac{\sin\theta}{(\theta-\cos\theta\, \sin\theta)^{1/3}}\,d\mu
    = \frac{3 \pi^{2/3}}{4},
    %\nonumber \\
    %&=&
    %\frac{3 \pi^{2/3}}{4}.
\end{eqnarray}
%\end{linenomath*}
and the constant $K_0$ is equal to
%\begin{linenomath*}
\begin{eqnarray}
    K_0 & = & 
    \frac{ 8\pi^2 j_0 R_{\ast}^3 }{L_0}
    \nonumber \\
    & = &
    1.35 \times 10^{5} \, 
    \left(\frac{C_{\ast}}{10^{10}}\right)
    \left(\frac{R_{\ast}}{10^{11}}\right)^{1/3}
    \left(\frac{B_{\ast}}{100~{\rm G}}\right)^{3/2}
    \times
    \nonumber \\
    & &
    \left(\frac{Z_{\rm i}^{2}}{\mu_{\rm e} \, \mu_{\rm i}}\right)^{-2/3}
    \left(\frac{n_{0}}{10^{13}}\right)^{-4/3} 
    ,
\end{eqnarray}
%\end{linenomath*}
Here, we have applied a fiducial scaling for the surface magnetic field $B_{\ast}$. We note that $K_0$ is  dependent on a number of parameters, especially considering that $n_0$ (eq.~[\ref{eq:n0}]) also depends on the stellar $\dot{M}$, $R_\ast$, and $v_\infty$. Thus a given $K_0$ may apply to a wide combination of observables. Table \ref{tab1} lists the values of $K_0$ considered for our model SEDs.  

The formal solution is then
%\begin{linenomath*}
\begin{equation}
\label{eq:formsol_hybridsed_norazin}
    \frac{L_\nu}{L_0} = 
    %\lambda^{-2/3}
     \left(\frac{\lambda}{\lambda_0}\right)^{-2/3} 
    + 
    1.4~K_0 \,\left(\frac{2}{\tau_0} \right)^{1/3} \, 
	%\lambda^{-1/6}
	 \left(\frac{\lambda}{\lambda_0}\right)^{-1/6} 
	,
\end{equation}
%\end{linenomath*}
While this result is essentially unphysical, since at sufficiently long wavelengths the luminosity integrated over wavelength is unbounded, (\ref{eq:formsol_hybridsed_norazin}) is useful for illustrating the transition of the SED from a thermal to a non-thermal spectrum. Model SEDs without the Razin suppression of the synchrotron emission (i.e., $a_0=0$) are shown as the red curves in Figure \ref{fig:hybrid}. These models show that the radio spectrum follows the curve expected from thermal emission at short wavelengths, but then becomes modified by the synchrotron emission
%(possibly dominated by the synchrotron) 
at long wavelengths.
%, and then eventually returns to a thermal spectrum.

Including the Razin effect (suppressing synchrotron emission) in the formal solution  yields 
\begin{eqnarray}
\frac{L_\nu}{L_0} & = & 
%\lambda^{-2/3}
 \left(\frac{\lambda}{\lambda_0}\right)^{-2/3} 
+ K_0 \,
%\lambda^{1/2}
 \left(\frac{\lambda}{\lambda_0}\right)^{1/2} 
\,\times \nonumber \\
 & &
 \int_{-1}^{1}\int_0^1
	u^{m-1/2} \, e^{-\tau_\nu(u,\mu)}\, e^{-\lambda/\lambda_R(u)}\,du\,d\mu
	,
\end{eqnarray}
where $\lambda_R (u) = \lambda_{\rm R}^{0} / u$ 
%\lambda_{\rm R}^{0} (r/R_{\ast})$ 
is the Razin wavelength (eqs.~[\ref{eq:synchB}] and [\ref{eq:razin_nu}]). The Razin wavelength scale constant is defined as
%\begin{linenomath*}
\begin{equation}
\label{eq:lam_r0}
    \lambda_{\rm R}^{0} = 
    0.015~{\rm cm} \,
    \frac{B_{\ast}}{100~{\rm G}}
    \frac{10^{13}}{n_0} .
\end{equation}
%\end{linenomath*}

Figure \ref{fig:hybrid} also shows model SEDs that include the Razin effect (black curves). The inclusion of Razin suppression significantly diminishes the synchrotron component at long wavelengths.

%\noindent Note that $\lambda$ is in centimeters for all preceding expressions.

%%%%%%%%%%%%%%%%%%%%%%%%%
\section{Discussion}	
\label{sec:conc}

The synthetic SEDs shown in Figure \ref{fig:hybrid} indicate that the inclusion of synchrotron emission can have a significant impact on the shape of the long-wavelength SED, even when Razin suppression is considered. In this regime, the SED is no longer dominated by thermal free-free emission, and so deviates from the canonical power law slope. This change is most pronounced when there is a fixed, constant ratio between the wind density and number density of relativistic electrons (left column).
%the number density of the relativistic electrons (parameterized by $C_{\gamma}$) is \christi{smaller (e.g., $C_{\gamma} \sim 1/r^2$, left column)}. 

For steeper distributions of relativistic electrons, the wavelength at which the free-free emission ceases to dominate the SED decreases. Razin suppression of the synchrotron emission becomes more pronounced, producing an SED with a slope that closely resembles the thermal power-law result. This implies that a sufficiently steep distribution of relativistic electrons may produce an SED with a power-law slope that mimics the result for thermal emission, particularly in a given waveband. 

Our results indicate that, in the absence of other factors, synchrotron emission can influence single-star SEDs, although a relatively fine degree of wavelength sampling may be needed for detection of the effect owing to the gradual change in SED slope. Although we do not directly consider magnetic massive stars as a part of this analysis, we briefly explore the effect of a latitudinally dependant toroidal field on the shape of the SED in Appendix \ref{app:toroid}.

%\christi{check radio instruments, what wavebands can they observe in?}
%\christi{Giant Metrewave Radio Telescope (GMRT): 21- 600cm}
%\christi{Atacama Large Millimeter/submillimeter Array (ALMA): 0.32 to 3.6 mm}
%\christi{Karl G. Jansky Very Large Array (VLA): 0.7 cm to 400 cm}

%\christi{This is a paragraph about what synch influence on single-star SED shape could mean for observations, and for refining the mass-loss rate calculations.}

The models reported here 
explore different distributions for the population of relativistic electrons using a power-law prescription for $C_{\gamma}$, although we offer no model for either how or where electrons are accelerated or transported throughout the wind. Indeed, the processes through which non-thermal synchrotron emission is produced in single, massive stars are still unclear. For stars with sufficient mass-loss rates to produce radio excesses in the wind, but with only modest surface magnetic fields (well below 1~kG), the radio photospheres is relatively extended compared to where synchrotron emission would typically form in the inner wind. As a result, little of the non-thermal component should escape to be observable. In order for non-thermal emission to compete in amplitude with the thermal component, electrons would need to be accelerated at relatively large radii of $\sim 10^1 - 10^2 R_\ast$.

Wind clumping may provide one pathway to achieving non-thermal emission in the extended stellar wind of a single massive star. The wind instabilities commonly understood to be associated with clumping give rise to shocks that are spread throughout the wind outflow. Such shocks have generally been associated with X-ray production \citep[e.g.,][]{1997A&A...322..167B,2011ApJS..194....7N}, and thus are a natural environment where electrons could be accelerated to relativistic energies \citep[e.g.,][]{1985ApJ...289..698W,1994Ap&SS.221..259C}. For example,
{\em XMM-Newton} and {\em Chandra} observations of the Wolf-Rayet star WR~6\footnote{\cite{2019A&A...624L...3S} have claimed that WR~6 may have a binary companion, which then could accounts for X-rays from large radii.  In an independent analysis, \cite{2020svos.conf..423S} have not been able to confirm the claim.  Moreover, the resolved X-ray line profile shapes observed by \cite{2015ApJ...815...29H} are consistent with predictions for a spherically symmetric terminal speed flow \citep{2001ApJ...549L.119I}.} shows evidence for X-ray emitting hot plasma emerging from $\sim 30R_\ast$ in the 
wind \citep{2012ApJ...747L..25O, 2013ApJ...775...29I, 2015ApJ...815...29H}, suggestive that shocks could form or persist to large radii.

More work will be needed to determine the role of wind instabilities in the production of non-thermal emission. Fully 3D simulations for the wind-driving instability mechanism 
%that is understood to spawn such stochastic structures (clumps) in massive stars winds 
have not yet been performed, although 2D simulations have been reported \citep[e.g.,][]{2003A&A...406L...1D,2018A&A...611A..17S}. A recent study by \citet{2018A&A...619A..59S} has suggested that porosity effects do not impact the radio photosphere of O-type stars. However, as \citet{2006A&A...452.1011V} point out, the results of 2D simulations inform our view of how shocks form and propagate outward in the wind. In particular, the 2D simulations indicate that round structures can form\footnote{Formally, these are rings in the 2D simulations; one may naturally expect spheroidal structures to form in fully 3D models}, which stands in opposition to the frequently invoked ``pancake'' shock geometry, referring to a shell that breaks up into multiple fragments of modest or small solid angle.

It is not yet clear how fully 3D simulations of the time-dependent wind flow would alter the expectations for particle acceleration. However, \citet{2016MNRAS.457.4123I} considered how porosity could impact the thermal radio emission in dense winds by including spherical clumps. Although a simplistic model in some respects, that study determined that porosity with spherical clumps implied that the radio photosphere forms deeper in the wind, as compared to the pure microclumping scenario. This may alleviate, to some extent, the distance to which relativistic electrons must survive or be accelerated in order for synchrotron emission to be significant.

The magnetospheres of single, magnetic massive stars provide another environment in which non-thermal emission can be produced. The confinement and channeling of the stellar wind by the magnetic field leads to the production of X-ray emission from shocks \citep[e.g.,][]{
    1997ApJ...485L..29B,
    2002ApJ...576..413U,
    2005ApJ...628..986G,
    2011AN....332..988O,
    2011MNRAS.416.1456O,
    2014ApJS..215...10N,
    2016AdSpR..58..680U}, 
which can accelerate electrons to the energies necessary for synchrotron emission. \citet{1997A&A...323..121B} used their model for {\it magnetically confined wind shocks} (MCWS) to  describe the X-ray emission of IQ Aur (HD 34452; A0p), and suggested the model could also be generally applied to the non-thermal radio emission from chemically peculiar A- and B-type (ApBp) stars. In this scenario, relativistic electrons are produced via a second-order Fermi acceleration mechanism, leading to the observed synchrotron emission. \citet{2004A&A...418..593T} reported an alternate model for magnetic chemically peculiar stars, suggesting that electrons are accelerated to relativistic speeds in current sheets that develop in the ``middle magnetosphere'' (where the stellar wind breaks open the magnetic field loops), while thermal-emitting plasma is trapped in the ``inner magnetosphere'' (the region within the closed magnetic field loops).

Recently, \citet{2021MNRAS.507.1979L} and \citet{2022arXiv220105512S} have proposed an alternate mechanism for the production of non-thermal emission in massive star magnetospheres. In contrast to the model developed by \citet{2004A&A...418..593T}, these authors argue for the production of synchrotron emission from a shellular ``radiation belt'' within the inner magnetosphere. The existance of such a structure appears to be supported by obervational evidence \citep[][]{2022arXiv220105512S}.

The theoretical framework for how electrons are accelerated to relativistic energies in single, massive stars will need to be refined. Yet there is  observational evidence for free-free and non-thermal emission in single, massive stars, particularly among the magnetic early-type star population. Thus, there is a pressing need for additional multiwavelength observations of radio SEDs, in order to further characterize this population. 

The models reported here can be used to help predict and interpret the behavior of radio SEDs from single, massive stars. In terms of the emissive model, future developments of this work would include exploring the effects of an aspherical wind density, plus consideration of pitch angle effects for the relativistic electrons. Also needed is further work for the acceleration of electrons to relativistic energies and their distribution throughout the wind.  These improvements are necessary to produce more quantitative predictions for SEDs involving both free-free and synchrotron contributions.  

%%%%%%%%%%%%%%%%%%%%%%%%%
\appendix

%%%%%%%%%%%%%%%%%%%%%%%%%

\section{Approximation of an Effective Photosphere for Synchrotron Emission}
\label{app:effphoto_synch}

We develop here an ``effective photosphere'' analysis for pure, optically thick synchrotron radiation, using the method outlined in \citet{1977ApJ...212..488C}.

The optical depth of a location in the wind is given by the integral of the synchrotron opacity (eq.~[\ref{eq:alphanu_p2}]) along the observer's line-of-sight. From Section \ref{sec:synch_emiss}, the field strength is described by eq.~(\ref{eq:synchB}), and we apply a constant power-law index of $p(r) = p_0 = 2$, so that the expression for the optical depth is written as
\begin{eqnarray}
\label{eq:tau_synch}
    \tau_\nu^{\rm s} & = & 
    \int (\kappa_\nu\rho)^{\rm s}\,dz,
    \nonumber \\
    & = & 
    \tau_0^{\rm s} \, R_{\ast}^{3+m}  
    %\lambda^3
     \left(\frac{\lambda}{\lambda_0}\right)^{3} 
    %\, 
    \int_{0}^{\theta_0(\varpi)}
    \frac{(\sin \theta)^{2+m}}{\varpi^{3+m}} d\theta,
\end{eqnarray}
where in the second equality above, we have applied the change of variables $z = \varpi \cot \theta$, where $\varpi$ is the impact parameter, and $\theta$ gives the angle between the radial direction and the observer's line-of-sight. For simplicity, we also define the quantity
%\begin{linenomath*}
\begin{equation}
    \label{eq:taus0}
    \tau_0^s = 
    5.1 \times 10^{-11} \,
    B_{\ast}^2 \,
    R_{\ast} \,
    \left(\frac{C_{\ast}}{10^{10}}\right). 
\end{equation}
%\end{linenomath*}
Under the assumption that $R_{\nu} \gg R_{\ast}$, $\theta_0 = \pi$, and the expression for the optical depth can be evaluated analytically,
%\begin{linenomath*}
\begin{equation}
    \label{eq:s_opd}
    \tau_\nu^{\rm s} = 
    \sqrt{\pi} \,
    \tau_0^{\rm s} \, 
    %\lambda^3 
     \left(\frac{\lambda}{\lambda_0}\right)^{3} 
    \,
    \left(\frac{R_{\ast}}{\varpi}\right)^{3+m}
    \frac{\Gamma(\frac{3+m}{2})}{\Gamma(\frac{4+m}{2})} .
\end{equation}
%\end{linenomath*}
For the choice of $m=0$, this yields
%integration of equation~(\ref{eq:tau_synch}) gives
%\begin{linenomath*}
\begin{equation}
    \label{eq:tausynch_m0}
    \tau_\nu^{\rm s} = 
    %\tau_0^{\rm s} \, \lambda^3 \,
    %\left(\frac{R_{\ast}}{\varpi}\right)^{3}
    %\left[ \frac{\theta_0}{2} - \frac{\sin(2 \theta_0)}{4} \right] 
    %=
    \frac{\pi}{2} \,
    \tau_0^{\rm s} \, 
    %\lambda^3 
     \left(\frac{\lambda}{\lambda_0}\right)^{3} 
    \,
    \left(\frac{R_{\ast}}{\varpi}\right)^{3}
    .
\end{equation}
%\end{linenomath*}
Using equations~\ref{eq:ff_opd} and \ref{eq:tau0} for the free-free optical depth, equations~\ref{eq:taus0} and \ref{eq:tausynch_m0} for the synchrotron optical depth, the fiducial values given throughout this paper, and a choice of $\theta = \pi/2$, we find $\tau_{\nu}^{\rm s}/\tau_\nu^{\rm ff}~\sim~10^{-4}$.

Equation~(\ref{eq:s_opd}) shows that the optical depth from pure synchrotron radiation increases more strongly with wavelength than the optical depth from free-free radiation (eq.~[\ref{eq:ff_opd}]) alone. As discussed in Section \ref{sec:synch_emiss}, the synchrotron radio photosphere will vary as $R_{\nu}^{\rm s} \propto \lambda^{3/(3+m)}$. For $m=0$, $R_{\nu}^{\rm s} \propto \lambda$, with a stronger dependence on wavelength than the free-free radio photosphere, which grows as $R_\nu \propto \lambda^{2/3}$ (eq.~[\ref{eq:rnu_ff}]). However, for $m=2$, the synchrotron effective radius of $R_{\nu}^{\rm s} \sim \lambda^{0.6}$ has a similar scaling to that for pure free-free radiation.

%%%%%%%%%%%%%%%%%%%%%%%%%

\section{Solution for a Latitudinal Dependence of the Toroidal Field}
\label{app:toroid}

%%%%%
\begin{figure}
\centering
\includegraphics[angle=0,width=0.55\columnwidth]{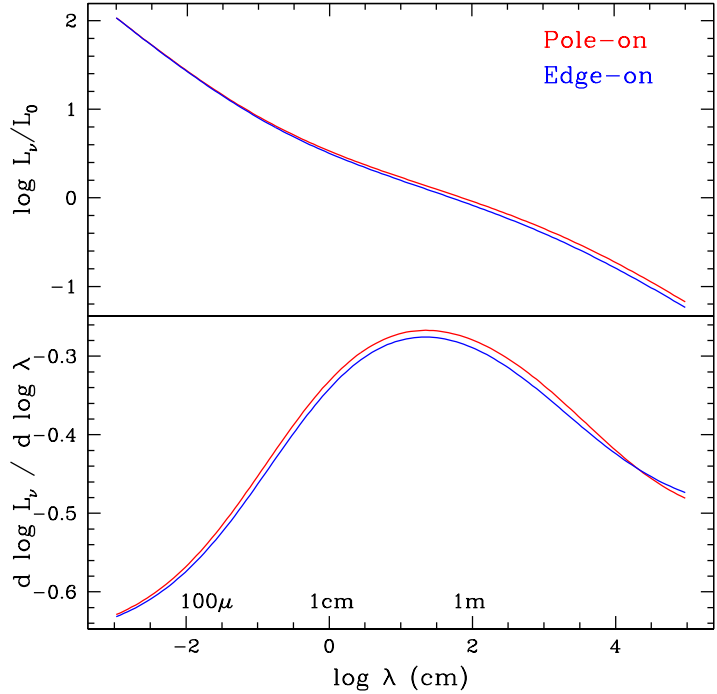}
\caption{
The luminosity of radio emission from both free-free and synchrotron processes, assuming a toroidal magnetic field with latitudinal dependence. The model assumes $K_0=200$, $m=0.5$, $a=3$. We show the resultant SED for inclinations of $0^\circ$ (a pole-on view, red curves) and $90^\circ$ (an edge-on view, blue curves).}
\label{fig:app} 
\end{figure}
%%%%%

There are three primary effects that can lead to deviations from
spherical symmetry in relation to the approach adopted in this
paper:  
\begin{enumerate}
\item[(1)] the magnetic field has both a latitudinal and an azimuthal dependence for strength and direction, 
\item[(2)] the pitch angles of the relativistic electrons are not random, and
\item[(3)] the wind density
itself is not spherically symmetric. 
\end{enumerate}
Here, we only address the first point. Although this condition likely implies the
third point, we assume the wind density is spherically symmetric, in order to isolate the effects of the field topology.

For synchrotron radiation formed in a wind that is thick to free-free opacity (Appendix~\ref{app:effphoto_synch}), at a radius far from the stellar surface, a toroidal component of the magnetic field is most
relevant, due to its slow decline with radius as $r^{-1}$. One may reasonably expect a latitudinal dependence of the field,
with

%\begin{linenomath*}
\begin{equation}
B_\varphi = \pm B_\ast\, \sin\vartheta\,\left(\frac{R_\ast}{r}\right)
	= \pm B_\ast\, \sin\vartheta\times u,
\end{equation}
%\end{linenomath*}
where $\pm B_\ast$ is the magnetic field strength at the stellar surface, and $\vartheta$ and $\varphi$ are spherical angular coordinates defined by the axis of symmetry for the toroidal field.

We define the observer's coordinates $(\theta,\alpha)$, with respect to the line-of-sight view that is
%viewing sightline
inclined by angle $i$ to the field symmetry axis. Thus $i=0^\circ$ gives a magnetic pole-on view, and $i=90^\circ$ gives an ``edge-on'' view of the magnetic equator. The
coordinate transformation between the observer's $(\theta,\alpha)$ and the latitude $\vartheta$ for the field is given by spherical 
trigonometry, with

%\begin{linenomath*}
\begin{equation}
\label{eq:cos_v}
\cos \vartheta = \cos\theta\,\cos i + \sin\theta\,\sin i\,\cos \alpha.
\end{equation}
%\end{linenomath*}

\noindent The free-free optical depth to a point $(r,\theta)$ in the
wind is given by equation~(\ref{eq:ff_opd}),
%\begin{linenomath*}
\begin{equation}
\tau_{\nu}^{\rm ff}(u,\theta) = \frac{\tau_0 \,u^3}{2}\,
%\lambda^2
 \left(\frac{\lambda}{\lambda_0}\right)^{2} 
\,\left(\frac
	{\theta - \cos\theta\,\sin\theta}{\sin^3\theta}\right) .
\end{equation}
%\end{linenomath*}

\noindent The Razin wavelength now becomes
%\begin{linenomath*}
\begin{equation}
    \lambda_R(u,\mu) =
    \lambda_{\rm R}^{0} \sin\vartheta \,
    u^{-1},
\end{equation}
%\end{linenomath*}
where the constant $\lambda_{\rm R}^{0}$ is given by equation~(\ref{eq:lam_r0}).

\noindent The luminosity of the radio emission, including both free-free and synchrotron processes, for a toroidal magnetic field with a latitudinal component is then

%\begin{linenomath*}
\begin{equation}
\frac{L_\nu}{L_0} = 
%\lambda^{-2/3}
 \left(\frac{\lambda}{\lambda_0}\right)^{-2/3} 
+ \frac{K_0}{2\pi}\,
%\lambda^{1/2}
 \left(\frac{\lambda}{\lambda_0}\right)^{1/2} 
\,\int
	u^{m-1/2}\,\left(\sin\vartheta\right)^{3/2}\, e^{-\tau_{\nu,{\rm ff}}}
	\, e^{-\lambda/\lambda_R}\,du\,d\mu\,d\alpha,
\end{equation}
%\end{linenomath*}

\noindent where $\vartheta = \vartheta(\mu,\alpha)$ and 
$\lambda_R = \lambda_R(u,\mu,\alpha)$. Evaluation of the integral requires the elimination of $\vartheta$ in
terms of $\theta$ and $\alpha$, using equation~(\ref{eq:cos_v}). The result will therefore depend on the viewing inclination, giving a luminosity that depends on~$i$. Figure~\ref{fig:app} shows results for calculations with $K_0=200$,
$m=0.5$, $a=3$, contrasting a pole-on view of $i=0^\circ$ (red curves) with an edge-on view of $i=90^\circ$ (blue curves). 

Figure~\ref{fig:app} reveals that including a latitudinal dependence of the toroidal field has fairly minimal impact on the SED shape or
the brightness level. However, if considered in conjunction with a non-spherical density, the effects could more substantial. An adjustment of the model to include a non-spherical density will be challenging. An axisymmetric density will introduce at least two more free parameters: a density contrast
and a distribution of density with latitude. Additionally, the free-free optical depth would no longer be generally analytic, so that a greater computational expense would be required to explore the range of outcomes with the multiple model parameters.  

%%%%%%%%%%%%%%%%%%%%%%%%%

\section{Applicability of Constant Expansion}
\label{app:appC}

%%%%%
\begin{figure}
\centering
\includegraphics[angle=0,width=0.55\columnwidth]{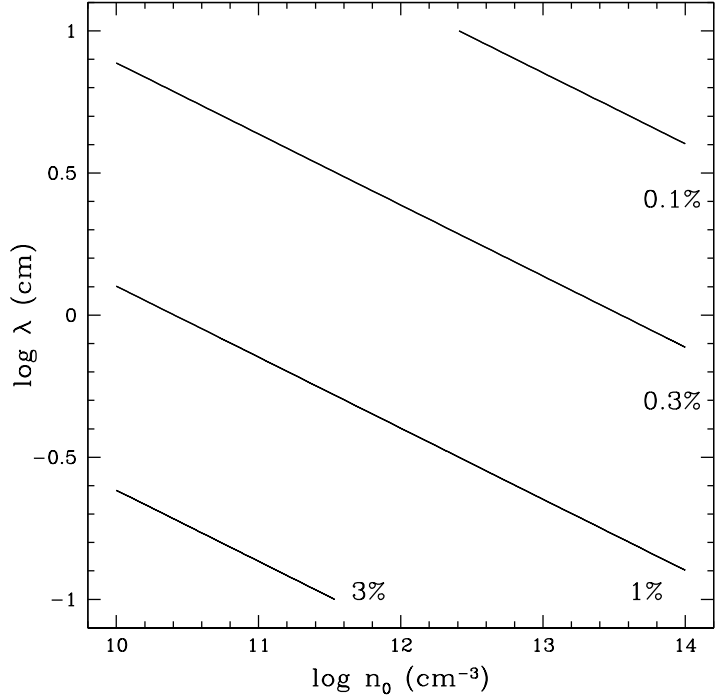}
\caption{Contour plot of the percent error in optical depth (eq.~[\ref{eq:err_opd}]). The axes are the wind density scale $n_0$ and wavelength~$\lambda$. }
\label{fig:appC} 
\end{figure}
%%%%%

This contribution has explored how synchrotron emission can alter the SED shape formed by a wind that is optically thick to thermal free-free opacity. As discussed above, high opacity results from a combination of dense winds and long wavelength radiation, and the canonical SED shape is a power law with $f_\nu \propto \lambda^{-2/3}$ (using the assumption for the Gaunt Factor of $g_\nu^{\rm ff} = 1$ adopted throughout this work). This result assumes a spherical, constantly expanding wind. Here, we comment on the applicability of this assumption.

Consider a canonical wind velocity law with $\beta = 1$, such that $v(r) = v_\infty\,(1-R_\ast/r)$. In normalized form, applying the substitution $u = R_{\ast}/r$, the velocity law can be expressed as $w(u) = 1-u$. The optical depth to any location $u$ in the wind (eq.~[\ref{eq:opd_ff}]), %accounting for the assumed wind velocity law, 
is then
    \begin{equation}
        \tau^{\rm ff} (u) = \tau_0 \left[\frac{1}{(1-u)} - (1-u) +2\ln(1-u)\right],
        \label{eq:vlawff}
    \end{equation}
with
    \begin{equation}
        \tau_0 = 1.99\times  10^8\,\left(\frac{\lambda}{\lambda_0}\right)^2\,\left(\frac{n_0}{10^{13}}\right)^2\,\left(\frac{R_\ast}{10^{11}}\right)\,g_\nu^{\rm ff}.
    \end{equation}
We note that while $\tau^{\rm ff}$ is an analytic function of $u$, the determinate of $u(\tau^{\rm ff})$ is implicit, requiring root finder methods for a solution.

For constant expansion, the characteristic radius of the free-free photosphere (eq.~[\ref{eq:rnu_ff}]) is then
    \begin{equation}
        u_\nu = \frac{R_\ast}{R_\nu} \equiv \left(\frac{\tau_0}{3}\right)^{1/3} ,
    \end{equation}
where the location $u_\nu$ has been defined by the condition of optical unity (eq.~[\ref{eq:tau_unity}]). 

One way to assess the applicability of our assumptions is to evaluate $\tau^{\rm ff}(u_\nu)$ using eq.~(\ref{eq:vlawff}). We define the relative error in the approximation for the optical depth as
    \begin{equation}
        \label{eq:err_opd}
        {\rm Error} = 100\% \times \left( \frac{\tau^{\rm ff}(u)-1}{\tau^{\rm ff}(u)} \right).
    \end{equation}
Fig.~(\ref{fig:appC}) provides a contour plot of this error. For simplicity, we assume a constant stellar luminosity $L_\ast$, and have adopted the scaling relations for hot massive star winds from \cite{2021arXiv210908164V}, with $\dot{M}~\propto~L_\ast^{2.2}\,M_\ast^{-1.3}\, T_\ast$, for stellar mass and effective temperature $M_\ast$ and $T_\ast$, respectively (see eq.~[4] of that paper). We assume for massive stars that $M_\ast~\propto~L_\ast$, and constant luminosity implies $T_\ast~\propto~R_\ast^{-1/2}$. For the wind terminal speed, $v_\infty \propto v_{\rm esc} \propto M_\ast^{1/2}\,R_\ast^{-1/2}$. We apply these scaling relations to eq.~(\ref{eq:n0}) to obtain the wind density scale $n_0$ in terms of these quantities,
    \begin{equation}
        n_0\propto \frac{\dot{M}}{R_\ast^2\,v_\infty}\propto \frac{L_\ast^{0.9}}{R_\ast^2\,M_\ast^{1/2}}
        \propto \frac{L_\ast^{0.4}}{R_\ast^2}.
    \end{equation}

Since the most luminous stars tend to have the highest wind mass-loss rates, the assumption of a constant $L_\ast$ implies $R_\ast \propto n_0^{1/2}$, and thus $\tau_0 \propto \lambda^2\,n_0^{3/2}$. This is the scaling relation used in Fig.~\ref{fig:appC}. While crude, the relation indicates that lower wind densities correspond to larger stars. For example, a wind density scale of $n_0 = 10^{13}$ cm$^{-3}$ and a stellar radius $R_\ast = 10^{11}~{\rm cm} \sim 1.5~R_{\odot}$ is appropriate to describe a WR star. In contrast, under our assumptions, a density scale of $n_0 = 10^{10}$ cm$^{-3}$ (appropriate for an O supergiant) would have a corresponding stellar radius of $R_\ast \sim 45~R_\odot$ (which in reality is about a factor of two too large).  
Overall, Fig.~\ref{fig:appC} reveals that across the span of wind density scales appropriate to stars ranging from O supergiants to WRs, at wavelengths from around 10~cm to 1~mm, the approximation of constant expansion is quite good. We find
%This evidenced by the fact that using the assumption to find $u_\nu$, then substituting back into the expression for optical depth that includes the wind velocity law, the 
a maximum error in the optical unity assumption of $\sim$5\%, for $\lambda=1$ mm with $n_0 =10^{10}$ cm$^{-3}$.

%%%%%%%%%%%%%%%%%%%%%%%%%
%%%%% END MATTER %%%%%

\acknowledgments
The authors wish to thank the anonymous referee for their careful reading and helpful comments, which have contributed to the final version of this paper.

C.E. and R.I. gratefully acknowledge that this material is based upon work supported by the National Science Foundation under Grant No. AST-2009412.

%The authors extend their thanks to the anonymous referee, whose helpful comments contributed to the improvement of this manuscript.

\bibliography{radio_sed_bib}

\begin{thebibliography}{}
\expandafter\ifx\csname natexlab\endcsname\relax\def\natexlab#1{#1}\fi
\providecommand{\url}[1]{\href{#1}{#1}}
\providecommand{\dodoi}[1]{doi:~\href{http://doi.org/#1}{\nolinkurl{#1}}}
\providecommand{\doeprint}[1]{\href{http://ascl.net/#1}{\nolinkurl{http://ascl.net/#1}}}
\providecommand{\doarXiv}[1]{\href{https://arxiv.org/abs/#1}{\nolinkurl{https://arxiv.org/abs/#1}}}

\bibitem[{{Abbott} {et~al.}(1986){Abbott}, {Beiging}, {Churchwell}, \&
  {Torres}}]{1986ApJ...303..239A}
{Abbott}, D.~C., {Beiging}, J.~H., {Churchwell}, E., \& {Torres}, A.~V. 1986,
  \apj, 303, 239, \dodoi{10.1086/164070}

\bibitem[{{Abbott} {et~al.}(1981){Abbott}, {Bieging}, \&
  {Churchwell}}]{1981ApJ...250..645A}
{Abbott}, D.~C., {Bieging}, J.~H., \& {Churchwell}, E. 1981, \apj, 250, 645,
  \dodoi{10.1086/159412}

\bibitem[{{Abbott} {et~al.}(1980){Abbott}, {Bieging}, {Churchwell}, \&
  {Cassinelli}}]{1980ApJ...238..196A}
{Abbott}, D.~C., {Bieging}, J.~H., {Churchwell}, E., \& {Cassinelli}, J.~P.
  1980, \apj, 238, 196, \dodoi{10.1086/157973}

\bibitem[{{Alecian} {et~al.}(2014){Alecian}, {Kochukhov}, {Petit}, {Grunhut},
  {Landstreet}, {Oksala}, {Wade}, {Hussain}, {Neiner}, {Bohlender}, \& {MiMeS
  Collaboration}}]{2014A&A...567A..28A}
{Alecian}, E., {Kochukhov}, O., {Petit}, V., {et~al.} 2014, \aap, 567, A28,
  \dodoi{10.1051/0004-6361/201323286}

\bibitem[{{Babel} \& {Montmerle}(1997{\natexlab{a}})}]{1997ApJ...485L..29B}
{Babel}, J., \& {Montmerle}, T. 1997{\natexlab{a}}, \apjl, 485, L29,
  \dodoi{10.1086/310806}

\bibitem[{{Babel} \& {Montmerle}(1997{\natexlab{b}})}]{1997A&A...323..121B}
---. 1997{\natexlab{b}}, \aap, 323, 121

\bibitem[{{Benaglia} {et~al.}(2015){Benaglia}, {Marcote}, {Mold{\'o}n},
  {Nelan}, {De Becker}, {Dougherty}, \& {Koribalski}}]{2015A&A...579A..99B}
{Benaglia}, P., {Marcote}, B., {Mold{\'o}n}, J., {et~al.} 2015, \aap, 579, A99,
  \dodoi{10.1051/0004-6361/201425595}

\bibitem[{{Berghoefer} {et~al.}(1997){Berghoefer}, {Schmitt}, {Danner}, \&
  {Cassinelli}}]{1997A&A...322..167B}
{Berghoefer}, T.~W., {Schmitt}, J.~H.~M.~M., {Danner}, R., \& {Cassinelli},
  J.~P. 1997, \aap, 322, 167

\bibitem[{{Bieging} {et~al.}(1989){Bieging}, {Abbott}, \&
  {Churchwell}}]{1989ApJ...340..518B}
{Bieging}, J.~H., {Abbott}, D.~C., \& {Churchwell}, E.~B. 1989, \apj, 340, 518,
  \dodoi{10.1086/167414}

\bibitem[{{Blomme} {et~al.}(2017){Blomme}, {Fenech}, {Prinja}, {Pittard}, \&
  {Morford}}]{2017A&A...608A..69B}
{Blomme}, R., {Fenech}, D.~M., {Prinja}, R.~K., {Pittard}, J.~M., \& {Morford},
  J.~C. 2017, \aap, 608, A69, \dodoi{10.1051/0004-6361/201731403}

\bibitem[{{Blomme} \& {Runacres}(1997)}]{1997A&A...323..886B}
{Blomme}, R., \& {Runacres}, M.~C. 1997, \aap, 323, 886

\bibitem[{{Bromm} \& {Larson}(2004)}]{2004ARA&A..42...79B}
{Bromm}, V., \& {Larson}, R.~B. 2004, \araa, 42, 79,
  \dodoi{10.1146/annurev.astro.42.053102.134034}

\bibitem[{{Brussaard} \& {van de Hulst}(1962)}]{1962RvMP...34..507B}
{Brussaard}, P.~J., \& {van de Hulst}, H.~C. 1962, Reviews of Modern Physics,
  34, 507, \dodoi{10.1103/RevModPhys.34.507}

\bibitem[{{Cassinelli} \& {Hartmann}(1977)}]{1977ApJ...212..488C}
{Cassinelli}, J.~P., \& {Hartmann}, L. 1977, \apj, 212, 488,
  \dodoi{10.1086/155068}

\bibitem[{{Chen} \& {White}(1991)}]{1991ApJ...366..512C}
{Chen}, W., \& {White}, R.~L. 1991, \apj, 366, 512, \dodoi{10.1086/169586}

\bibitem[{{Chen} \& {White}(1994)}]{1994Ap&SS.221..259C}
---. 1994, \apss, 221, 259, \dodoi{10.1007/BF01091158}

\bibitem[{{Daley-Yates} {et~al.}(2019){Daley-Yates}, {Stevens}, \&
  {ud-Doula}}]{2019MNRAS.489.3251D}
{Daley-Yates}, S., {Stevens}, I.~R., \& {ud-Doula}, A. 2019, \mnras, 489, 3251,
  \dodoi{10.1093/mnras/stz1982}

\bibitem[{{De Becker}(2018)}]{2018BSRSL..87..185D}
{De Becker}, M. 2018, Bulletin de la Societe Royale des Sciences de Liege, 87,
  185

\bibitem[{{De Becker} {et~al.}(2017){De Becker}, {Benaglia}, {Romero}, \&
  {Peri}}]{2017A&A...600A..47D}
{De Becker}, M., {Benaglia}, P., {Romero}, G.~E., \& {Peri}, C.~S. 2017, \aap,
  600, A47, \dodoi{10.1051/0004-6361/201629110}

\bibitem[{{Dessart} \& {Owocki}(2003)}]{2003A&A...406L...1D}
{Dessart}, L., \& {Owocki}, S.~P. 2003, \aap, 406, L1,
  \dodoi{10.1051/0004-6361:20030810}

\bibitem[{{Donati} \& {Landstreet}(2009)}]{2009ARA&A..47..333D}
{Donati}, J.-F., \& {Landstreet}, J.~D. 2009, \araa, 47, 333,
  \dodoi{10.1146/annurev-astro-082708-101833}

\bibitem[{{Donati} {et~al.}(2006){Donati}, {Howarth}, {Jardine}, {Petit},
  {Catala}, {Landstreet}, {Bouret}, {Alecian}, {Barnes}, {Forveille},
  {Paletou}, \& {Manset}}]{Donati2006b}
{Donati}, J.~F., {Howarth}, I.~D., {Jardine}, M.~M., {et~al.} 2006, \mnras,
  370, 629, \dodoi{10.1111/j.1365-2966.2006.10558.x}

\bibitem[{{Dougherty} {et~al.}(2003){Dougherty}, {Pittard}, {Kasian}, {Coker},
  {Williams}, \& {Lloyd}}]{2003A&A...409..217D}
{Dougherty}, S.~M., {Pittard}, J.~M., {Kasian}, L., {et~al.} 2003, \aap, 409,
  217, \dodoi{10.1051/0004-6361:20031048}

\bibitem[{{Dougherty} \& {Williams}(2000)}]{2000MNRAS.319.1005D}
{Dougherty}, S.~M., \& {Williams}, P.~M. 2000, \mnras, 319, 1005,
  \dodoi{10.1046/j.1365-8711.2000.03837.x}

\bibitem[{{Eichler} \& {Usov}(1993)}]{1993ApJ...402..271E}
{Eichler}, D., \& {Usov}, V. 1993, \apj, 402, 271, \dodoi{10.1086/172130}

\bibitem[{{Ellison} \& {Eichler}(1985)}]{1985PhRvL..55.2735E}
{Ellison}, D.~C., \& {Eichler}, D. 1985, \prl, 55, 2735,
  \dodoi{10.1103/PhysRevLett.55.2735}

\bibitem[{{Falceta-Gon{\c c}alves} \& {Abraham}(2012)}]{2012MNRAS.423.1562F}
{Falceta-Gon{\c c}alves}, D., \& {Abraham}, Z. 2012, \mnras, 423, 1562,
  \dodoi{10.1111/j.1365-2966.2012.20978.x}

\bibitem[{{Fossati} {et~al.}(2015){Fossati}, {Castro}, {Sch{\"o}ller},
  {Hubrig}, {Langer}, {Morel}, {Briquet}, {Herrero}, {Przybilla}, {Sana},
  {Schneider}, {de Koter}, \& {BOB Collaboration}}]{2015A&A...582A..45F}
{Fossati}, L., {Castro}, N., {Sch{\"o}ller}, M., {et~al.} 2015, \aap, 582, A45,
  \dodoi{10.1051/0004-6361/201526725}

\bibitem[{{Fullerton} {et~al.}(2006){Fullerton}, {Massa}, \&
  {Prinja}}]{2006ApJ...637.1025F}
{Fullerton}, A.~W., {Massa}, D.~L., \& {Prinja}, R.~K. 2006, \apj, 637, 1025,
  \dodoi{10.1086/498560}

\bibitem[{{Gagn{\'e}} {et~al.}(2005){Gagn{\'e}}, {Oksala}, {Cohen}, {Tonnesen},
  {ud-Doula}, {Owocki}, {Townsend}, \& {MacFarlane}}]{2005ApJ...628..986G}
{Gagn{\'e}}, M., {Oksala}, M.~E., {Cohen}, D.~H., {et~al.} 2005, \apj, 628,
  986, \dodoi{10.1086/430873}

\bibitem[{{Grunhut} {et~al.}(2017){Grunhut}, {Wade}, {Neiner}, {Oksala},
  {Petit}, {Alecian}, {Bohlender}, {Bouret}, {Henrichs}, {Hussain},
  {Kochukhov}, \& {MiMeS Collaboration}}]{2017MNRAS.465.2432G}
{Grunhut}, J.~H., {Wade}, G.~A., {Neiner}, C., {et~al.} 2017, \mnras, 465,
  2432, \dodoi{10.1093/mnras/stw2743}

\bibitem[{{Heger} {et~al.}(2003){Heger}, {Fryer}, {Woosley}, {Langer}, \&
  {Hartmann}}]{2003ApJ...591..288H}
{Heger}, A., {Fryer}, C.~L., {Woosley}, S.~E., {Langer}, N., \& {Hartmann},
  D.~H. 2003, \apj, 591, 288, \dodoi{10.1086/375341}

\bibitem[{{Heger} \& {Woosley}(2010)}]{2010ApJ...724..341H}
{Heger}, A., \& {Woosley}, S.~E. 2010, \apj, 724, 341,
  \dodoi{10.1088/0004-637X/724/1/341}

\bibitem[{{Hopkins} {et~al.}(2012){Hopkins}, {Quataert}, \&
  {Murray}}]{2012MNRAS.421.3522H}
{Hopkins}, P.~F., {Quataert}, E., \& {Murray}, N. 2012, \mnras, 421, 3522,
  \dodoi{10.1111/j.1365-2966.2012.20593.x}

\bibitem[{{Huenemoerder} {et~al.}(2015){Huenemoerder}, {Gayley}, {Hamann},
  {Ignace}, {Nichols}, {Oskinova}, {Pollock}, {Schulz}, \&
  {Shenar}}]{2015ApJ...815...29H}
{Huenemoerder}, D.~P., {Gayley}, K.~G., {Hamann}, W.~R., {et~al.} 2015, \apj,
  815, 29, \dodoi{10.1088/0004-637X/815/1/29}

\bibitem[{{Hummer}(1988)}]{1988ApJ...327..477H}
{Hummer}, D.~G. 1988, \apj, 327, 477, \dodoi{10.1086/166210}

\bibitem[{{Ignace}(2001)}]{2001ApJ...549L.119I}
{Ignace}, R. 2001, \apjl, 549, L119, \dodoi{10.1086/319141}

\bibitem[{{Ignace}(2016)}]{2016MNRAS.457.4123I}
---. 2016, \mnras, 457, 4123, \dodoi{10.1093/mnras/stw216}

\bibitem[{{Ignace} {et~al.}(1998){Ignace}, {Cassinelli}, \&
  {Bjorkman}}]{1998ApJ...505..910I}
{Ignace}, R., {Cassinelli}, J.~P., \& {Bjorkman}, J.~E. 1998, \apj, 505, 910,
  \dodoi{10.1086/306189}

\bibitem[{{Ignace} {et~al.}(2013){Ignace}, {Gayley}, {Hamann}, {Huenemoerder},
  {Oskinova}, {Pollock}, \& {McFall}}]{2013ApJ...775...29I}
{Ignace}, R., {Gayley}, K.~G., {Hamann}, W.~R., {et~al.} 2013, \apj, 775, 29,
  \dodoi{10.1088/0004-637X/775/1/29}

\bibitem[{{Ignace} {et~al.}(2003){Ignace}, {Quigley}, \&
  {Cassinelli}}]{2003ApJ...596..538I}
{Ignace}, R., {Quigley}, M.~F., \& {Cassinelli}, J.~P. 2003, \apj, 596, 538,
  \dodoi{10.1086/377597}

\bibitem[{{Klement} {et~al.}(2017){Klement}, {Carciofi}, {Rivinius},
  {Matthews}, {Vieira}, {Ignace}, {Bjorkman}, {Mota}, {Faes}, {Bratcher},
  {Cur{\'e}}, \& {{\v S}tefl}}]{2017A&A...601A..74K}
{Klement}, R., {Carciofi}, A.~C., {Rivinius}, T., {et~al.} 2017, \aap, 601,
  A74, \dodoi{10.1051/0004-6361/201629932}

\bibitem[{{Kochukhov} {et~al.}(2011){Kochukhov}, {Lundin}, {Romanyuk}, \&
  {Kudryavtsev}}]{Kochukhov2011}
{Kochukhov}, O., {Lundin}, A., {Romanyuk}, I., \& {Kudryavtsev}, D. 2011, \apj,
  726, 24, \dodoi{10.1088/0004-637X/726/1/24}

\bibitem[{{Kurapati} {et~al.}(2017){Kurapati}, {Chandra}, {Wade}, {Cohen},
  {David-Uraz}, {Gagne}, {Grunhut}, {Oksala}, {Petit}, {Shultz}, {Sundqvist},
  {Townsend}, \& {ud-Doula}}]{2017MNRAS.465.2160K}
{Kurapati}, S., {Chandra}, P., {Wade}, G., {et~al.} 2017, \mnras, 465, 2160,
  \dodoi{10.1093/mnras/stw2838}

\bibitem[{{Lamers} \& {Cassinelli}(1996)}]{1996ASPC...98..162L}
{Lamers}, H.~J.~G.~L.~M., \& {Cassinelli}, I.~P. 1996, in Astronomical Society
  of the Pacific Conference Series, Vol.~98, From Stars to Galaxies: the Impact
  of Stellar Physics on Galaxy Evolution, ed. C.~{Leitherer},
  U.~{Fritze-von-Alvensleben}, \& J.~{Huchra}, 162

\bibitem[{{Langer}(2012)}]{2012ARA&A..50..107L}
{Langer}, N. 2012, \araa, 50, 107, \dodoi{10.1146/annurev-astro-081811-125534}

\bibitem[{{Leitherer} {et~al.}(1995){Leitherer}, {Chapman}, \&
  {Koribalski}}]{1995ApJ...450..289L}
{Leitherer}, C., {Chapman}, J.~M., \& {Koribalski}, B. 1995, \apj, 450, 289,
  \dodoi{10.1086/176140}

\bibitem[{{Leitherer} {et~al.}(1997){Leitherer}, {Chapman}, \&
  {Koribalski}}]{1997ApJ...481..898L}
---. 1997, \apj, 481, 898, \dodoi{10.1086/304096}

\bibitem[{{Leto} {et~al.}(2006){Leto}, {Trigilio}, {Buemi}, {Umana}, \&
  {Leone}}]{2006A&A...458..831L}
{Leto}, P., {Trigilio}, C., {Buemi}, C.~S., {Umana}, G., \& {Leone}, F. 2006,
  \aap, 458, 831, \dodoi{10.1051/0004-6361:20054511}

\bibitem[{{Leto} {et~al.}(2017){Leto}, {Trigilio}, {Oskinova}, {Ignace},
  {Buemi}, {Umana}, {Ingallinera}, {Todt}, \& {Leone}}]{2017MNRAS.467.2820L}
{Leto}, P., {Trigilio}, C., {Oskinova}, L., {et~al.} 2017, \mnras, 467, 2820,
  \dodoi{10.1093/mnras/stx267}

\bibitem[{{Leto} {et~al.}(2018){Leto}, {Trigilio}, {Oskinova}, {Ignace},
  {Buemi}, {Umana}, {Ingallinera}, {Leone}, {Phillips}, {Agliozzo}, {Todt}, \&
  {Cerrigone}}]{2018MNRAS.476..562L}
{Leto}, P., {Trigilio}, C., {Oskinova}, L.~M., {et~al.} 2018, \mnras, 476, 562,
  \dodoi{10.1093/mnras/sty244}

\bibitem[{{Leto} {et~al.}(2021){Leto}, {Trigilio}, {Krti{\v{c}}ka}, {Fossati},
  {Ignace}, {Shultz}, {Buemi}, {Cerrigone}, {Umana}, {Ingallinera}, {Bordiu},
  {Pillitteri}, {Bufano}, {Oskinova}, {Agliozzo}, {Cavallaro}, {Riggi}, {Loru},
  {Todt}, {Giarrusso}, {Phillips}, {Robrade}, \& {Leone}}]{2021MNRAS.507.1979L}
{Leto}, P., {Trigilio}, C., {Krti{\v{c}}ka}, J., {et~al.} 2021, \mnras, 507,
  1979, \dodoi{10.1093/mnras/stab2168}

\bibitem[{{Lucy} \& {White}(1980)}]{1980ApJ...241..300L}
{Lucy}, L.~B., \& {White}, R.~L. 1980, \apj, 241, 300, \dodoi{10.1086/158342}

\bibitem[{{Madau} \& {Dickinson}(2014)}]{2014ARA&A..52..415M}
{Madau}, P., \& {Dickinson}, M. 2014, \araa, 52, 415,
  \dodoi{10.1146/annurev-astro-081811-125615}

\bibitem[{{Morel} {et~al.}(2015){Morel}, {Castro}, {Fossati}, {Hubrig},
  {Langer}, {Przybilla}, {Sch{\"o}ller}, {Carroll}, {Ilyin}, {Irrgang},
  {Oskinova}, {Schneider}, {D{\'\i}az}, {Briquet}, {Gonz{\'a}lez},
  {Kharchenko}, {Nieva}, {Scholz}, {de Koter}, {Hamann}, {Herrero}, {Ma{\'\i}z
  Apell{\'a}niz}, {Sana}, {Arlt}, {Barb{\'a}}, {Dufton}, {Kholtygin}, {Mathys},
  {Piskunov}, {Reisenegger}, {Spruit}, \& {Yoon}}]{2015IAUS..307..342M}
{Morel}, T., {Castro}, N., {Fossati}, L., {et~al.} 2015, in New Windows on
  Massive Stars, ed. G.~{Meynet}, C.~{Georgy}, J.~{Groh}, \& P.~{Stee}, Vol.
  307, 342--347, \dodoi{10.1017/S1743921314007054}

\bibitem[{{Naz{\'e}} {et~al.}(2014){Naz{\'e}}, {Petit}, {Rinbrand}, {Cohen},
  {Owocki}, {ud-Doula}, \& {Wade}}]{2014ApJS..215...10N}
{Naz{\'e}}, Y., {Petit}, V., {Rinbrand}, M., {et~al.} 2014, \apjs, 215, 10,
  \dodoi{10.1088/0067-0049/215/1/10}

\bibitem[{{Naz{\'e}} {et~al.}(2011){Naz{\'e}}, {Broos}, {Oskinova}, {Townsley},
  {Cohen}, {Corcoran}, {Evans}, {Gagn{\'e}}, {Moffat}, {Pittard}, {Rauw},
  {ud-Doula}, \& {Walborn}}]{2011ApJS..194....7N}
{Naz{\'e}}, Y., {Broos}, P.~S., {Oskinova}, L., {et~al.} 2011, \apjs, 194, 7,
  \dodoi{10.1088/0067-0049/194/1/7}

\bibitem[{{Nugis} {et~al.}(1998){Nugis}, {Crowther}, \&
  {Willis}}]{1998A&A...333..956N}
{Nugis}, T., {Crowther}, P.~A., \& {Willis}, A.~J. 1998, \aap, 333, 956

\bibitem[{{Oskinova} {et~al.}(2012){Oskinova}, {Gayley}, {Hamann},
  {Huenemoerder}, {Ignace}, \& {Pollock}}]{2012ApJ...747L..25O}
{Oskinova}, L.~M., {Gayley}, K.~G., {Hamann}, W.~R., {et~al.} 2012, \apjl, 747,
  L25, \dodoi{10.1088/2041-8205/747/2/L25}

\bibitem[{{Oskinova} {et~al.}(2011{\natexlab{a}}){Oskinova}, {Hamann},
  {Cassinelli}, {Brown}, \& {Todt}}]{2011AN....332..988O}
{Oskinova}, L.~M., {Hamann}, W.-R., {Cassinelli}, J.~P., {Brown}, J.~C., \&
  {Todt}, H. 2011{\natexlab{a}}, Astronomische Nachrichten, 332, 988,
  \dodoi{10.1002/asna.201111602}

\bibitem[{{Oskinova} {et~al.}(2011{\natexlab{b}}){Oskinova}, {Todt}, {Ignace},
  {Brown}, {Cassinelli}, \& {Hamann}}]{2011MNRAS.416.1456O}
{Oskinova}, L.~M., {Todt}, H., {Ignace}, R., {et~al.} 2011{\natexlab{b}},
  \mnras, 416, 1456, \dodoi{10.1111/j.1365-2966.2011.19143.x}

\bibitem[{{Owocki} {et~al.}(1988){Owocki}, {Castor}, \&
  {Rybicki}}]{1988ApJ...335..914O}
{Owocki}, S.~P., {Castor}, J.~I., \& {Rybicki}, G.~B. 1988, \apj, 335, 914,
  \dodoi{10.1086/166977}

\bibitem[{{Owocki} \& {ud-Doula}(2004)}]{2004ApJ...600.1004O}
{Owocki}, S.~P., \& {ud-Doula}, A. 2004, \apj, 600, 1004,
  \dodoi{10.1086/380123}

\bibitem[{{Panagia} \& {Felli}(1975)}]{1975A&A....39....1P}
{Panagia}, N., \& {Felli}, M. 1975, \aap, 39, 1

\bibitem[{{Parkin} {et~al.}(2014){Parkin}, {Pittard}, {Naz{\'e}}, \&
  {Blomme}}]{2014A&A...570A..10P}
{Parkin}, E.~R., {Pittard}, J.~M., {Naz{\'e}}, Y., \& {Blomme}, R. 2014, \aap,
  570, A10, \dodoi{10.1051/0004-6361/201423833}

\bibitem[{{Pittard} {et~al.}(2006){Pittard}, {Dougherty}, {Coker}, {O'Connor},
  \& {Bolingbroke}}]{2006A&A...446.1001P}
{Pittard}, J.~M., {Dougherty}, S.~M., {Coker}, R.~F., {O'Connor}, E., \&
  {Bolingbroke}, N.~J. 2006, \aap, 446, 1001,
  \dodoi{10.1051/0004-6361:20053649}

\bibitem[{{Pittard} {et~al.}(2021){Pittard}, {Romero}, \&
  {Vila}}]{2021MNRAS.504.4204P}
{Pittard}, J.~M., {Romero}, G.~E., \& {Vila}, G.~S. 2021, \mnras, 504, 4204,
  \dodoi{10.1093/mnras/stab1107}

\bibitem[{{Puls} {et~al.}(2006){Puls}, {Markova}, {Scuderi}, {Stanghellini},
  {Taranova}, {Burnley}, \& {Howarth}}]{2006A&A...454..625P}
{Puls}, J., {Markova}, N., {Scuderi}, S., {et~al.} 2006, \aap, 454, 625,
  \dodoi{10.1051/0004-6361:20065073}

\bibitem[{{Rybicki} \& {Lightman}(1986)}]{1986rpa..book.....R}
{Rybicki}, G.~B., \& {Lightman}, A.~P. 1986, {Radiative Processes in
  Astrophysics} (Wiley-VCH), 400

\bibitem[{{Schmid-Burgk}(1982)}]{1982A&A...108..169S}
{Schmid-Burgk}, J. 1982, \aap, 108, 169

\bibitem[{{Schmutz} \& {Koenigsberger}(2019)}]{2019A&A...624L...3S}
{Schmutz}, W., \& {Koenigsberger}, G. 2019, \aap, 624, L3,
  \dodoi{10.1051/0004-6361/201935094}

\bibitem[{{Shultz} {et~al.}(2022){Shultz}, {Owocki}, {ud-Doula}, {Biswas},
  {Bohlender}, {Chandra}, {Das}, {David-Uraz}, {Khalack}, {Kochukhov},
  {Landstreet}, {Leto}, {Monin}, {Neiner}, {Rivinius}, \&
  {Wade}}]{2022arXiv220105512S}
{Shultz}, M.~E., {Owocki}, S.~P., {ud-Doula}, A., {et~al.} 2022, arXiv
  e-prints, arXiv:2201.05512.
\newblock \doarXiv{2201.05512}

\bibitem[{{Silvester} {et~al.}(2014){Silvester}, {Kochukhov}, \&
  {Wade}}]{Silvester2014}
{Silvester}, J., {Kochukhov}, O., \& {Wade}, G.~A. 2014, \mnras, 440, 182,
  \dodoi{10.1093/mnras/stu306}

\bibitem[{{St-Louis} {et~al.}(2020){St-Louis}, {Moffat}, {Ramiaramanantsoa}, \&
  {Lenoir-Craig}}]{2020svos.conf..423S}
{St-Louis}, N., {Moffat}, A.~F.~J., {Ramiaramanantsoa}, T., \& {Lenoir-Craig},
  G. 2020, in Stars and their Variability Observed from Space, ed. C.~{Neiner},
  W.~W. {Weiss}, D.~{Baade}, R.~E. {Griffin}, C.~C. {Lovekin}, \& A.~F.~J.
  {Moffat}, 423--426

\bibitem[{{Sundqvist} {et~al.}(2018){Sundqvist}, {Owocki}, \&
  {Puls}}]{2018A&A...611A..17S}
{Sundqvist}, J.~O., {Owocki}, S.~P., \& {Puls}, J. 2018, \aap, 611, A17,
  \dodoi{10.1051/0004-6361/201731718}

\bibitem[{{Sundqvist} \& {Puls}(2018)}]{2018A&A...619A..59S}
{Sundqvist}, J.~O., \& {Puls}, J. 2018, \aap, 619, A59,
  \dodoi{10.1051/0004-6361/201832993}

\bibitem[{{Trigilio} {et~al.}(2004){Trigilio}, {Leto}, {Umana}, {Leone}, \&
  {Buemi}}]{2004A&A...418..593T}
{Trigilio}, C., {Leto}, P., {Umana}, G., {Leone}, F., \& {Buemi}, C.~S. 2004,
  \aap, 418, 593, \dodoi{10.1051/0004-6361:20040060}

\bibitem[{{ud-Doula} \& {Naz{\'e}}(2016)}]{2016AdSpR..58..680U}
{ud-Doula}, A., \& {Naz{\'e}}, Y. 2016, Advances in Space Research, 58, 680,
  \dodoi{10.1016/j.asr.2015.09.025}

\bibitem[{{ud-Doula} \& {Owocki}(2002)}]{2002ApJ...576..413U}
{ud-Doula}, A., \& {Owocki}, S.~P. 2002, \apj, 576, 413, \dodoi{10.1086/341543}

\bibitem[{{Van Loo} {et~al.}(2004){Van Loo}, {Runacres}, \&
  {Blomme}}]{2004A&A...418..717V}
{Van Loo}, S., {Runacres}, M.~C., \& {Blomme}, R. 2004, \aap, 418, 717,
  \dodoi{10.1051/0004-6361:20034480}

\bibitem[{{Van Loo} {et~al.}(2005){Van Loo}, {Runacres}, \&
  {Blomme}}]{2005A&A...433..313V}
---. 2005, \aap, 433, 313, \dodoi{10.1051/0004-6361:20041973}

\bibitem[{{Van Loo} {et~al.}(2006){Van Loo}, {Runacres}, \&
  {Blomme}}]{2006A&A...452.1011V}
---. 2006, \aap, 452, 1011, \dodoi{10.1051/0004-6361:20054266}

\bibitem[{{Vink}(2021)}]{2021arXiv210908164V}
{Vink}, J.~S. 2021, arXiv e-prints, arXiv:2109.08164.
\newblock \doarXiv{2109.08164}

\bibitem[{{Wade} {et~al.}(2016){Wade}, {Neiner}, {Alecian}, {Grunhut}, {Petit},
  {Batz}, {Bohlender}, {Cohen}, {Henrichs}, {Kochukhov}, {Landstreet},
  {Manset}, {Martins}, {Mathis}, {Oksala}, {Owocki}, {Rivinius}, {Shultz},
  {Sundqvist}, {Townsend}, {ud-Doula}, {Bouret}, {Braithwaite}, {Briquet},
  {Carciofi}, {David-Uraz}, {Folsom}, {Fullerton}, {Leroy}, {Marcolino},
  {Moffat}, {Naz{\'e}}, {Louis}, {Auri{\`e}re}, {Bagnulo}, {Bailey},
  {Barb{\'a}}, {Blaz{\`e}re}, {B{\"o}hm}, {Catala}, {Donati}, {Ferrario},
  {Harrington}, {Howarth}, {Ignace}, {Kaper}, {L{\"u}ftinger}, {Prinja},
  {Vink}, {Weiss}, \& {Yakunin}}]{2016MNRAS.456....2W}
{Wade}, G.~A., {Neiner}, C., {Alecian}, E., {et~al.} 2016, \mnras, 456, 2,
  \dodoi{10.1093/mnras/stv2568}

\bibitem[{{White}(1985)}]{1985ApJ...289..698W}
{White}, R.~L. 1985, \apj, 289, 698, \dodoi{10.1086/162933}

\bibitem[{{White} \& {Chen}(1992)}]{1992ApJ...387L..81W}
{White}, R.~L., \& {Chen}, W. 1992, \apjl, 387, L81, \dodoi{10.1086/186310}

\bibitem[{{White} \& {Chen}(1994)}]{1994Ap&SS.221..295W}
---. 1994, \apss, 221, 295, \dodoi{10.1007/BF01091161}

\bibitem[{{Williams} {et~al.}(1994){Williams}, {van der Hucht}, \&
  {Spoelstra}}]{1994A&A...291..805W}
{Williams}, P.~M., {van der Hucht}, K.~A., \& {Spoelstra}, T.~A.~T. 1994, \aap,
  291, 805

\bibitem[{{Woosley} {et~al.}(2002){Woosley}, {Heger}, \&
  {Weaver}}]{2002RvMP...74.1015W}
{Woosley}, S.~E., {Heger}, A., \& {Weaver}, T.~A. 2002, Reviews of Modern
  Physics, 74, 1015, \dodoi{10.1103/RevModPhys.74.1015}

\bibitem[{{Wright} \& {Barlow}(1975)}]{1975MNRAS.170...41W}
{Wright}, A.~E., \& {Barlow}, M.~J. 1975, \mnras, 170, 41,
  \dodoi{10.1093/mnras/170.1.41}

\end{thebibliography}

\end{document}